\documentclass[10pt,journal,compsoc]{IEEEtran}

\newif\ifsubmission
\newif\ifanonymous
\submissiontrue
\anonymousfalse
\ifdefined\FORSUBMISSION \submissiontrue \fi
\ifdefined\FORANONYMOUS \anonymoustrue \fi

\ifCLASSOPTIONcompsoc
  \usepackage[nocompress]{cite}
\else
  \usepackage{cite}
\fi
\usepackage[T1]{fontenc}
\usepackage{booktabs}
\usepackage{graphicx}
\usepackage{pdfpages}    %
\ifdefined\pdfinclusioncopyfonts \pdfinclusioncopyfonts=1 \fi
\input{glyphtounicode-cmex}
\usepackage{amsmath}
\usepackage{amssymb}
\usepackage{url}
\usepackage{xcolor}
\usepackage{pgfplots}    %
\pgfplotsset{compat=1.18}
\usepackage{xr-hyper}    %
\usepackage[hidelinks]{hyperref}
\space %
\usepackage[capitalise,noabbrev]{cleveref}

\ifsubmission
  
\else
  
\fi

\newif\ifhascite
\AtBeginDocument{%
  \let\gccfOrigCite\cite \renewcommand{\cite}{\global\hascitetrue\gccfOrigCite}%
  \let\gccfOrigNocite\nocite \renewcommand{\nocite}{\global\hascitetrue\gccfOrigNocite}}

\begin{document}
\bstctlcite{IEEEtran:BSTcontrol}

\title{Operating-Point Failure of Frozen Guardrail Probes Under Backbone Updates}

\ifanonymous
  \author{Anonymous authors}
\else
  \author{Yusuf~Khalid~Shire and Sang-Chul~Kim%
    \IEEEcompsocitemizethanks{\IEEEcompsocthanksitem Yusuf Khalid Shire and Sang-Chul Kim are with the School of Computer Science, Kookmin University, Seoul 02707, Republic of Korea.\protect\\ 
      E-mail: khalidshire@kookmin.ac.kr, sckim7@kookmin.ac.kr
    \IEEEcompsocthanksitem Corresponding authors: Yusuf Khalid Shire (khalidshire@kookmin.ac.kr); Sang-Chul Kim (sckim7@kookmin.ac.kr).}}
\fi

\markboth{IEEE Transactions on Dependable and Secure Computing}%
{Operating-Point Failure of Frozen Guardrail Probes Under Backbone Updates}

\IEEEtitleabstractindextext{%
\begin{abstract}
We freeze a guardrail probe and threshold and measure its operating point in 18 cells of 16 public language-model updates. 
At a 1\% target the threshold fails closed in 8 of 9 cells of a first phase and closed or open in 3 of 9 of a registered second phase (descriptive); the phases also differ in update kind and input format. 
The largest rise is 1.01\% to 15.12\%, while no such cell's area under the receiver operating characteristic curve moves more than 0.0058; a benchmark's staleness criteria call none stale. 
A rule from them flags 0.019 of layer cells whose rate doubles or halves; a label-free rate estimate flags 0.918, clearing 0.898 of the rest. 
A two-moment model matches the change within a factor of 1.5 in 0.905 of 63 registered cells, 0.773 of the 22 that moved. 
Re-setting the threshold on stored benign prompts restores the false-positive rate without new labels; poisoning 1\% of them leaves at most 0.017 of harmful prompts flagged. 
An attacker controlling the update cut detection of one of three targeted categories in three of five seeds, unseen by both checks; a post hoc labelled check flagged it. 
\space%
\end{abstract}
\begin{IEEEkeywords}
Large language models, guardrails, activation probes, operating point, model updates, threshold recalibration.
\end{IEEEkeywords}}

\AddToShipoutPictureFG*{\AtPageLowerLeft{\put(\LenToUnit{0.5\paperwidth},\LenToUnit{13pt}){\makebox(0,0)[b]{%
  \parbox[b]{\textwidth}{\centering\footnotesize This work has been submitted to the IEEE for possible publication.
  Copyright may be transferred without notice, after which this version may no longer be accessible.}}}}}
\maketitle
\IEEEdisplaynontitleabstractindextext
\IEEEpeerreviewmaketitle

\IEEEraisesectionheading{\section{Introduction}\label{sec:introduction}}

\IEEEPARstart{T}{he} deployer of a guardrail probe sets its threshold once, to flag 1\% of benign prompts, and freezes it; the backbone whose hidden states the probe reads is then replaced by a release that the deployer did not schedule. 
The probe is a linear classifier on those states, and it flags a prompt whose score reaches the threshold $T$.
After the update the probe and $T$ stay as deployed, and the probe reads activations it was not fitted to.
If the update moves the scores of benign prompts, the share of them at or above $T$ moves too.
When the share at least doubles, the guardrail fails closed and flags benign prompts it passed before; when it falls to half or less, the guardrail fails open, more permissive than it was set to be. 
Either way, the deployer still runs a control that they believe flags 1\% of benign prompts. 

Duan~\cite{duan2026monitors} reports that fine-tuning often leaves a frozen activation probe stale, judged by criteria on AUROC, and also records recall at 1\% and 5\% false-positive rate. 
Duan gives no false-positive rate realised at a frozen threshold.
Whether the operating point, the false-positive and true-positive rates at $T$, survives an update is a different question, and an AUROC-based check does not answer it.
AUROC asks whether harmful prompts still score above benign ones, and it reads no threshold.
An update that moves every score by the same amount leaves AUROC unchanged and can still carry benign prompts across $T$.
Varshney~\cite{varshney2026cosine} shows this on constructed scores: a translation leaves AUROC unchanged while accuracy at a fixed threshold falls, and re-fitting the threshold recovers most of the accuracy lost to a translation but a minority of that lost to an attenuation of the signal.
Luo et al.~\cite{luo2026measuring} report that a threshold frozen on internal harmfulness scores can lose more F1 than the score loses AUROC as the inputs shift.
Biswas~\cite{biswas2026confidently} freezes prompt-injection detectors' thresholds at a 1\% false-positive rate on clean benign prompts and reports the realised rates as the attacks shift; the detectors' scores are poorly calibrated on attacks before any shift as well as after, which a pooled calibration error can hide. 
Luo and Biswas move the inputs, while we hold the prompts fixed and replace the backbone.
Concurrent work does too: one submission measures the false-positive rate realised at a threshold set for a 5\% target across a model's post-training checkpoints and re-sets that threshold on benign inputs~\cite{anon2026basecase}, and Tsatsos~\cite{tsatsos2026probes} judges base-model probes reusable on their successors by AUC (Section~\ref{sec:related}). 
We hold a 1\% threshold, include released version increments, register a model of the change before scoring it, score a label-free check as a detector, and measure what poisoning costs the label-free repairs and what an attacker who controls the update can do. 

We measure the operating point before and after backbone updates.
Our headline set has 18 cells at the probe's selected layer, all on pairs of public checkpoints: seven Phase~1 pairs, two of them repeated on a second dataset, and nine Phase~2 pairs that a registered rule fixed before any Phase~2 evaluation score existed. 
Each cell has 5 probe seeds and 1000 re-partitions of its benign prompts into the part that sets $T$ and the part that tests it. 
Beside the headline set we fine-tune one model on three corpora at three doses and test repairs that need no new labels. 
We attack both the update and a repair's reference, and test a canary check against the update attack (Section~\ref{sec:res-attack}).

\begin{enumerate}
\item A two-moment model, which takes the mean and standard deviation of the same benign prompts' scores after the update, accounts for the change in the false-positive rate within $\times1.5$ in 0.905 of the 63 Phase~2 cells whose predictions we registered before any Phase~2 evaluation score existed (HELD; Section~\ref{sec:res-mech}). 
\item The estimated-rate check, which reads only the moments of the stored benign prompts' scores, flags 0.918 of the damaged Phase~1 layer cells and clears 0.898 of the undamaged ones (descriptive), where the AUC rule, built from two of Duan's criteria, flags 0.019 of the damaged cells; a Kolmogorov--Smirnov (KS) test on the same scores flags more of the damaged cells but clears only 0.543 of the undamaged ones (HELD; Section~\ref{sec:res-mech}). 
\item At the 1\% target the frozen threshold fails closed in 8 of the 9 Phase~1 headline cells and in 2 of the 9 Phase~2 cells, and fails open in one Phase~2 cell; at the largest rise its false-positive rate goes from 1.01\% to 15.12\%, while no headline cell's AUROC changes by more than 0.0058 and Duan's criteria call none of the 18 cells stale (descriptive; Section~\ref{sec:res-op}). 
The two phases also differ in the kind of update and in how the model reads the prompt, so the difference between their counts cannot be credited to the kind of update (Section~\ref{sec:res-op}).
\item Re-setting the threshold on stored or fresh benign prompts needs no new labels and returns the false-positive rate to its target in all nine Phase~1 cells (descriptive; Section~\ref{sec:res-repair}). 
Poisoning the reference with the harmful prompts that the frozen probe scores highest after the update costs the re-set 10 points of true-positive rate at 0.5\% or 1\% of a 1000-item reference, the only size we poisoned, and at 1\% the re-set flags 0.006 to 0.017 of harmful prompts; quantile transfer on 1\% traffic loses 10 points at 1\%, where it still flags 0.153 to 0.489, and flags 0.009 to 0.018 at 2\% (descriptive). 
\item Realignment, fitted on the stored calibration prompts, keeps AUROC by a criterion we extended from Duan's, which the frozen probe already met (HELD), but leaves the false-positive rate at $T$ outside $\times1.5$ of its target in 9 of 9 units with the ridge map in each of Phase~1, Phase~2 and our fine-tunes, and in 6, 4 and 3 of 9 with the Procrustes map; we had registered at least 6 of 9, which HELD for the ridge map and for Procrustes in Phase~1 and FAILED for Procrustes in Phase~2 and our fine-tunes, where fewer units stayed outside (Section~\ref{sec:res-repair}). 
\item An attacker who controls the update can move the operating point on one of three targeted harm categories, in three of its five seeds, while the AUC rule and the estimated-rate check stay silent, refuting our registered hypothesis that no single attacker evades both (Hypothesis~A2, REFUTED); a post hoc canary on held-out labelled harmful prompts of all three categories flags each such cell (Section~\ref{sec:res-attack}). 
\item We label results as registered, with their verdicts, or as descriptive, and report all 137 entries of our registry, failures included: the 107 that state a prediction and the 30 outcome-only or descriptive ones, which state none (Supplement~\ref{supp:registry}). 
\end{enumerate}

Table~\ref{tab:claims} gives each of our claims, except those on the attacks and on our registry, with its primary evidence and its status.
\begin{table*}[t]
\caption{Our claims, their primary evidence and their status. The full registry, every row with its verdict, is Table~\mbox{\ref{tab:registry}}.}
\label{tab:claims}
\centering\scriptsize\setlength{\tabcolsep}{3pt}
\begin{tabular}{@{}lp{0.45\textwidth}p{0.18\textwidth}p{0.29\textwidth}@{}}
\toprule
ID & Claim & Primary evidence & Status \tabularnewline
\midrule
C1 & \raggedright\textit{Damage.} Benign released updates move the frozen threshold's false-positive rate (the paired ratio) while AUROC changes little; the AUC rule misses most failures. & \raggedright Table~\ref{tab:headline} & \raggedright No registered primary row; descriptive.\newline 6 rows: 4 held, 1 did not hold, 1 no verdict. \tabularnewline
C2 & \raggedright\textit{Mechanism.} The two-moment model of the probe's score distribution explains the change (it uses the post-update moments of the same evaluation negatives whose false-positive rate it gives), and the label-free estimated-rate check is an out-of-sample check: its moments come from the calibration negatives only. & \raggedright The change accounted for within $\times1.5$ in 0.905 of 63 cells (Section~\ref{sec:res-mech}) & \raggedright Rows~77 to 79, HELD.\newline 17 rows: 8 held, 5 did not hold, 4 no verdict. \tabularnewline
C3 & \raggedright\textit{Intent.} At matched dose, safety-oriented fine-tunes move the operating point more than general ones. & \raggedright Table~\ref{tab:lora} & \raggedright Row~20, HELD.\newline 8 rows: 5 held, 2 did not hold, 1 no verdict. \tabularnewline
C4 & \raggedright\textit{Repairs.} Threshold-level recalibration on stored or fresh benign references returns the false-positive rate to its target; restandardisation is valid only with a class-matched reference; repairs estimated from live traffic are biased or exposed to prevalence drift and poisoning. & \raggedright Mean false-positive rate 0.96 to 1.02 times its target after the re-set on stored or fresh benign prompts (Section~\ref{sec:res-repair}) & \raggedright Row~50, DESCRIPTIVE.\newline 23 rows: 12 held, 8 did not hold, 3 no verdict. \tabularnewline
C5 & \raggedright\textit{Health check.} A check that estimates the false-positive rate at the deployed threshold, from a trusted set of benign prompts, detects most damaging updates; a check on ranking misses most of them, and a check on generic distribution shift detects most of them but also alarms on about half of the harmless ones. & \raggedright Table~\ref{tab:fourc}; Fig.~\ref{fig:fourc-scatter}; the KS test, Section~\ref{sec:res-mech} & \raggedright Row~34, OUTCOME ONLY.\newline 7 rows: 3 held, 2 did not hold, 2 no verdict. \tabularnewline
\bottomrule
\end{tabular}
\par\smallskip\parbox{\textwidth}{\scriptsize Claim: as worded in our claims record, with the paper's names for the AUC rule, the estimated-rate check and the false-positive rate. Primary evidence: the table or figure of the claim's primary result, or its key number and the section that reports it, then any further result the wording states. Status: that result's registry row and verdict, or, where it has no registry row, so stated; then every registry row that our claims record assigns to a part of the claim, except parts the record has retired, once each, so the count can include rows on parts the one-line wording does not state, counted as held (HELD), did not hold (FAILED, NOT SUPPORTED, REFUTED) or no verdict (OUTCOME ONLY, DESCRIPTIVE, UNDEFINED, INCONCLUSIVE, WITHDRAWN BEFORE RUNNING).}
\end{table*}
\space%
\space%
\section{Related Work}\label{sec:related}

\textit{Activation probes as guardrails.}
An activation probe is a classifier trained on a language model's hidden states, a method Belinkov~\cite{belinkov2022probing} reviews; as a guardrail, it flags a prompt whose score reaches a threshold.
Sahoo et al.~\cite{sahoo2026robust} report that simulated drift of the embeddings a toxicity classifier reads cuts its AUROC to chance while its confidence stays high.
On one model, Goldowsky-Dill et al.~\cite{goldowskydill2025deception} report AUROC above 0.998 against control responses while a threshold at 1\% false-positive rate on them misses 4\% of deceptive responses on one dataset, and McKenzie et al.~\cite{mckenzie2025highstakes} compare probes for high-stakes interactions with language-model baselines at that rate. 
Gao~\cite{gao2026alarm} measures the detection and false-alarm rates that activation monitors realise at thresholds locked on a calibration split, on a test split read by the same frozen model; we read the same prompts before and after an update. 
A detector can instead read the input text: Llama Guard~\cite{inan2023llamaguard} classifies prompts and responses with a fine-tuned language model, and PIDS-Bench~\cite{shire2026pidsbench} evaluates text-level prompt-injection detectors on hard-benign, obfuscated and distribution-shifted inputs and reports their false-positive rate and F1 at fixed thresholds and across a full threshold sweep.

\textit{Probes under model change.}
Fine-tuning on benign datasets such as Alpaca, one of our three corpora, can degrade a model's safety alignment~\cite{qi2024finetuning}.
Duan~\cite{duan2026monitors} asks whether activation monitors survive model updates, under quantisation, LoRA~\cite{hu2022lora}, merged LoRA and QLoRA.
There, a label-free activation realignment fitted on paired unlabelled activations of the base and updated models repaired all 111 repair-relevant stale cells of a 216-cell targeted grid, with none needing score calibration, where a repair succeeds if it recovers at least half the lost AUROC or restores AUROC above 0.70. 
A concurrent submission~\cite{anon2026drift} carries frozen probes across the supervised, preference and reinforcement-learning stages of OLMo-3 7B and Qwen3-4B, reports AUROC and the true-positive rate at 1\% false-positive rate against a probe refit on each checkpoint, finds most of the gap after the supervised stage, and finds that label-free representation alignment closes much of it but can plateau below the refit or fall below the frozen probe; our realignment results (Section~\ref{sec:res-repair}) read instead the false-positive rate at the frozen threshold. 
Duan's predictor estimates the change in AUROC, or whether that change is a big drop, from features observable before the update is evaluated. 
Our estimated-rate check (Section~\ref{sec:twomoment}) estimates the false-positive-rate ratio after the update. 
Bakker and Louizos~\cite{bakker2026relative} freeze probes across quantisation, LoRA and QLoRA updates of Qwen2.5, and global anchored probes reduce the AUROC change they report (Section~\ref{sec:disc-lim}).
Under compression, a concurrent submission~\cite{anon2026frozenharm} finds that a frozen input-harm probe's retained AUROC does not certify that the compressed model still refuses, and that the probe's flag rate at a fixed threshold falls under 2:4 pruning. 
Das and Fioretto~\cite{das2026neurofilter} apply probes trained on Qwen2.5-7B to Qwen2.5-7B-Instruct, report diminished test accuracy at most layers, and advise retraining probes after fine-tuning and recalibrating them after model updates; on that step, our pair 1e (Section~\ref{sec:pairs}), the frozen threshold's false-positive rate stays within the two-fold cut (Table~\ref{tab:headline}).
Wilhelm et al.~\cite{wilhelm2026monitoring} carry a monitor across fine-tuned adapters, treated as unseen models, and report F1.
Their threshold is a constant at 0.5; ours is set to a target false-positive rate. 
Concurrent with this work, an anonymous submission to ACL Rolling Review, public since 13 October 2026~\cite{anon2026basecase}, trains probes on the base and post-training checkpoints of OLMo-3-7B and Apertus-8B and evaluates each, without retraining, on the other checkpoints and on fine-tunes of the final one. 
It holds a threshold set for a 5\% false-positive rate on the source checkpoint, reports that rankings are preserved while that threshold drifts, and finds that updates of an already post-trained checkpoint move it less than updates of the base model. 
It re-sets the threshold by matching the old threshold's decisions on benign inputs, which restores the rate only when those inputs come from the population the threshold was validated on, and recommends revalidating a probe at its operating point after every update.
Our frozen-threshold measurement, re-set and recommendation are of the same kind; ours hold a 1\% target, cover released version increments as well as post-training steps, and measure what poisoning the re-set's stored reference costs (Section~\ref{sec:res-repair}). 
Tsatsos~\cite{tsatsos2026probes} applies base-model probes, with their standardisation, unchanged to the instruction-tuned and distilled successors of four base models, on raw text as our Phase~1 does, and judges them reusable by AUC; at the probes' default threshold, which is not set for a target rate, the transferred probes flag more benign prompts that sound harmful on the distilled successors, and the angle between two models' safety directions is a poor guide to whether a probe still works.

\textit{Attacks.}
A model fine-tuned with a rank-one LoRA can learn to edit its activations along the monitored direction and lower a monitor's true-positive rate, and monitors refit on its clean activations stay evadable~\cite{keenan2026evade}.
Our attacks A1m and A2 (Section~\ref{sec:attacks}) use this mechanism, delivered by the party that produces the update: A2 shifts one harm category's scores under a KL term that keeps the model near a clean update and is scored against the checks a deployer can run without labels.
Fine-tuning can teach a model to evade monitors it did not see in training when a trigger is present~\cite{mcguinness2025chameleons}, and under reinforcement learning token-level monitors are highly vulnerable while max-pooling and attention-based probes remain robust~\cite{gupta2025rlobfuscation}.
Data poisoning has been demonstrated against anomaly detectors~\cite{rubinstein2009antidote,kloft2012centroid}, test-time adaptation~\cite{wu2023tta,cong2024tepa}, conformal prediction~\cite{li2024conformal} and the update loop of jailbreak-detection classifiers~\cite{huang2026rapidpoison}.
Kloft and Laskov~\cite{kloft2012centroid} derive the optimal poisoning of an online centroid anomaly detector and bound the attacker's effort, including under a bounded false-positive rate; for a threshold re-set on a poisoned reference, the analogous count is an order statistic of the reference, which we registered in closed form (Section~\ref{sec:res-repair}).
In test-time adaptation, malicious samples in a test batch can change the predictions on its benign samples through the batch's normalisation statistics~\cite{wu2023tta}.
This paper poisons a repair's reference set (Section~\ref{sec:attacks}).

\textit{Recalibration under distribution shift.}
Berlin~\cite{berlin2026calibration} notes that retraining changes a security detector's scores and, at each model release, refits a calibration of the whole benign false-positive-rate curve on that model's benign scores, so that a calibrated threshold keeps its target rate across releases.
Our threshold re-set (Section~\ref{sec:repairs}) does this for one threshold, on stored benign prompts after a backbone update; our quantile transfer instead carries $T$'s rank among the scores of unlabelled traffic, not benign samples, across the update.
Schirmer et al.~\cite{schirmer2026online} bound a monitor's false alarms by risk control, with calibration data drawn exchangeably from the deployment distribution, and Tong et al.~\cite{tong2018neyman} bound a classifier's false-positive rate with high probability from an i.i.d. sample; scores from before an update need not satisfy either assumption, while a re-set on the post-update scores of benign prompts drawn from the deployment's population could, and would then carry the finite-sample guarantee that our plug-in re-set lacks.
Under label shift only the class prior changes; Lipton et al.~\cite{lipton2018labelshift} detect and correct such a shift with a black-box predictor, and Saerens et al.~\cite{saerens2002adjusting} adjust a classifier's outputs to new prior probabilities.
Here the class prior enters through unlabelled traffic, the reference of quantile transfer and one restandardisation variant.
For probes that predict a vision-language model's errors, a concurrent submission~\cite{anon2026rankings} finds that post-training mostly translates the probe's scores, so that rankings survive while probability calibration drifts, re-centres the scores on unlabelled inputs under a common shift, and shows that no map computed from unlabelled scores alone can be calibrated for every update.
Our two-moment model needs the change in spread as well as the shift (Section~\ref{sec:res-mech}), and our label-free repairs target the false-positive rate, which depends on benign scores alone.
AdaBN~\cite{li2018adabn} re-estimates batch-normalisation statistics on unlabelled data from a new input distribution, with the other weights unchanged.
The same re-estimation improves robustness to image corruptions~\cite{schneider2020covariate} and calibration under covariate shift~\cite{nado2020predictiontime}.
Our restandardisation applies the same principle outside the network, to the standardisation statistics of the probe's input (Section~\ref{sec:repairs}), and a concurrent submission~\cite{anon2026frozenharm} re-centres a frozen harm probe's inputs on a compressed model's mean activations in the same way, with the probe's weights fixed. %
\space%
\section{Setting and Threat Model}\label{sec:setting}

\textit{Deployment.}
Guardrails can be stacked around one model \cite{alotaibi2026layered}; we study a single one, a linear probe trained once on a backbone's hidden states, with its threshold $T$ frozen as Section~\ref{sec:introduction} describes.
Kram\'ar et al.~\cite{kramar2026production} state that their findings informed the deployment of misuse-mitigation probes in user-facing instances of Gemini, and evaluate probes at a single threshold chosen on validation data.
A backbone replaced under a frozen probe and threshold is an assumption of our study.

\textit{What the deployer holds.}
After an update the deployer holds the new weights and can score any prompt through them.
The deployer also holds the stored calibration set: the benign prompts that set $T$, with their pre-update scores.
The deployer sees unlabelled traffic and gets no new labels.
We assume that the deployer knows neither the update's training data nor its arrival time.
For the AUROC-based check, we also assume that the deployer keeps the labelled prompts they held before the update; re-scoring them needs no new labels.

\textit{Checks that need no new labels.}
The AUROC-based check applies the three criteria of Duan~\cite{duan2026monitors} to the AUROC of the labelled prompts the deployer kept, before and after the update (Section~\ref{sec:metrics}).
Representation similarity, such as linear CKA~\cite{kornblith2019cka}, compares the old and new models' hidden states on the same prompts.
It needs no labels and reads neither the probe nor $T$.
Re-measurement at the operating point scores stored or fresh benign prompts, the references of Section~\ref{sec:repairs}, through the new model and counts those at or above $T$.
This measures the false-positive rate at $T$ on those prompts.
The true-positive rate at $T$ needs labelled harmful prompts that the deployer already holds.

\textit{The attacker.}
The attacker knows the frozen probe, its layer and its threshold.
They hold one of three positions: the producer of the update controls its weights; a contributor to live traffic controls a share of the prompts on which quantile transfer and restandardisation on traffic are fitted; and a supplier of the stored reference, or an insider who can write to it, controls a share of the prompts on which the threshold is re-set.
With the update, they move the frozen scores on every prompt or on one harmful category only, inside the stealth bound of Section~\ref{sec:attacks}, under which an attack on one category is a lower bound for its family.
With a reference, the attacker fills their share with existing prompts chosen by their scores; the stored prompts come from public datasets, so such an attacker can read them.
Out of scope are adaptive attackers, including those who know the canary prompts (held-out labelled harmful prompts that, in a deployment, the deployer alone holds; ours are public-dataset prompts chosen by a rule our repository releases, Section~\ref{sec:attacks}), and attackers who write new prompt text, such as a trigger phrase.

\textit{What we do not claim.}
We make no claim about any specific deployed system.
Both checkpoints of each headline pair are public (Section~\ref{sec:pairs}); our LoRA fine-tunes, attacked models, probes and thresholds are our own.
We assume the attacker's access; we do not show that any party has it.
Nor do we claim that an attacker can choose when an update arrives or place an attack in an update window.
\space%
\begin{table}[t]
\caption{Configuration changes of the update pairs: the fields of config.json, at the pinned revisions, that differ from the pre- to the post-update checkpoint, among the RoPE base, RoPE scaling, sliding window, maximum positions, hidden size, layer count and vocabulary size.}
\label{tab:config}
\centering\footnotesize
\begin{tabular}{@{}p{0.26\columnwidth}p{0.68\columnwidth}@{}}
\toprule
Pairs & Fields that change (pre$\to$post) \\
\midrule
1a--1d, 1f, 2c, 2e--2i & none \\
1e & max positions 131072$\to$32768 \\
1g, 2b & RoPE scaling none$\to$llama3; max positions 8192$\to$131072 \\
1h, 2a & RoPE base $10^{4}\to10^{6}$; sliding window 4096$\to$none \\
2d & RoPE scaling longrope, long\_factor and short\_factor changed \\
2j & vocabulary 32000$\to$32768 \\
\bottomrule
\end{tabular}
\end{table}
\space      %
\section{Method}\label{sec:method}

\subsection{Probe, layer and threshold}\label{sec:probe}

For a prompt $x$, let $h_\ell(x)$ be the hidden state at layer $\ell$, averaged over tokens.
In Phase~1 we feed both models the raw prompt, with no chat template, and average over every token the attention mask keeps, so the input format is fixed across the update. 
In Phase~2 each instruct model gets its released chat template, and we average over the user-message tokens only; the last templated position is a sensitivity arm. 
The one Phase~2 base checkpoint, OLMo-2-1124-7B, gets raw text, and its SFT successor its chat template. 

The probe is an $\ell_2$-regularised logistic regression on features standardised by the mean and standard deviation of the prompts it is fitted on, taken on the pre-update model; a standard deviation below $10^{-8}$ is set to~1. 
We minimise $\tfrac12\lVert w\rVert^2 + C\sum_i \mathcal{L}_{\log}(x_i,y_i)$ over the weight vector $w$, with an unpenalised intercept $b$ (scikit-learn, L-BFGS, at most 2000 iterations). 
The guardrail score is $s(x)=w^\top z(x)+b$, where $z(x)$ is the standardised $h_\ell(x)$ at the selected layer.

Each probe is trained on 4000 prompts (2000 harmful, 2000 benign). 
For each seed $s\in\{0,\dots,4\}$ we permute this set, hold out the first 20\%, and fit on the remaining 3200 prompts. 
We select $C$ and the layer $L^*$ jointly, $C$ from $\{10^{-4},3\cdot10^{-4},10^{-3},3\cdot10^{-3},10^{-2},3\cdot10^{-2},10^{-1},1\}$ and $L^*$ from every second layer, by the mean held-out AUROC over the five seeds, on the pre-update model only. 
The selected $C$ was $10^{-3}$ or $3\cdot10^{-3}$ for every pre-update model; Supplement~\ref{supp:revisions} gives each cell's $L^*$ and $C$. 
We report $C=1$, the library default, as a sensitivity arm. 
A cell is one pair on one dataset at one layer: Phase~1 evaluates every second layer, and Phase~2 evaluates $L^*$ and a fixed grid of six layers. 
The headline set is the 18 cells at $L^*$. 

The threshold $T$ is the smallest distinct calibration-negative score $t$ such that at most 1\% of calibration negatives score at or above $t$ ($T=+\infty$ if no such score exists). 
Phase~1 calibrates on 1000 negatives and Phase~2 on 5000. 
A 5\% threshold is a registered sensitivity arm. 

After an update we keep $w$, $b$, the standardisation statistics, $L^*$ and $T$, and apply them to the post-update model's activations.
Nothing is refit.
This frozen guardrail is the object of study; Section~\ref{sec:repairs} relaxes it.

In Phase~1, at 1\% on 1000 negatives, $T$ sits near the tenth-highest negative score; in Phase~2, on 5000, it sits near the fiftieth. 
The operating point thus depends on which negatives a split puts in the calibration part.
Our estimator therefore re-partitions the pooled calibration and evaluation negatives 1000 times at their original sizes, in both phases, re-sets $T$ in each draw, and averages over draws; only the study of reference pools, the attacks and the ridge realignment use fewer, 200, and the Procrustes realignment 50. 
The positives and the probes are the same in every draw. 
Bands are the 2.5th and 97.5th percentiles over draws. 
An interval pooled over pairs instead comes from a bootstrap that resamples the seven to nine pairs it pools (5000 replicates); the pairs share checkpoints, so such an interval need not hold 95\% coverage. 
We report the fixed split alongside.

The probe-training prompts, calibration negatives and evaluation negatives are disjoint by item and by exact text in both phases. 
Phase~1 drops duplicate texts and then splits each class from one seeded shuffle; Phase~2 puts each near-duplicate cluster whole in one split. 
We checked this on the frozen files by item identifier and text hash; 8 item pairs across these splits match only after case and whitespace normalisation, 3 of them between probe training and evaluation, and none in the Phase~1 WildGuardMix splits (Supplement~\ref{supp:disclosures}). 

\subsection{Update pairs}\label{sec:pairs}

An update pair $(P,Q)$ is a pre-update and a post-update checkpoint with the same parameter count and hidden size, each pinned to a revision hash (Supplement~\ref{supp:revisions}). 
For our 18 numbered pairs, an \emph{update} is the released weights together with their released configuration and, in Phase~2, their released format, so it can change more than the weights. 
Probe, layer and threshold are set on $P$ and evaluated on $Q$.
This rule removes Gemma-2$\to$Gemma-3, whose hidden size differs. 
We exclude successive Llama Guard releases as primary pairs because they changed the hazard taxonomy, not only the representation. 
Supplement~\ref{supp:unrun} lists candidate pairs classed clean but not run.
We number the pairs 1a to 1h and 2a to 2j in the order below; Supplement~\ref{supp:revisions} lists each pair's checkpoints and cells at $L^*$. 
Seven of these 18 numbered pairs change the model configuration; Table~\ref{tab:config} gives each change, and 2a, the one fail-open headline cell, changes the RoPE base and drops the sliding window. 

\textit{Phase~1} has seven headline pairs, plus 1h (Supplement~\ref{supp:extra}). 
Six are base$\to$instruct releases: Mistral-7B v0.1, Gemma-2-9B, Llama-3-8B, Llama-3.1-8B, Qwen2.5-7B and OLMo-2-1124-7B. 
The seventh is a base-model version update, Llama-3-8B$\to$Llama-3.1-8B. 
These pairs predate our registered selection rules, and no surviving written rule yields them. 
Probes use Aegis~2.0~\cite{ghosh2025aegis2}; the Mistral and Llama-3 base$\to$instruct pairs are repeated on WildGuardMix~\cite{han2024wildguard}, giving nine cells at $L^*$. 
Each evaluation split holds 2000 negatives and 2000 positives. 
Outside the headline we run one Mistral-7B v0.1 checkpoint on V100 and A100 GPUs with the same weights, the \emph{hardware control}, to isolate the shift from hardware. 
We also run LoRA~\cite{hu2022lora} fine-tunes of that checkpoint on Alpaca~\cite{taori2023alpaca}, HH-harmless~\cite{bai2022hhrlhf} and PKU-SafeRLHF~\cite{ji2025pkusaferlhf} for 25, 100 and 400 steps. 
The fine-tunes use rank 16, $\alpha=32$, dropout 0.05, learning rate $2\cdot10^{-4}$, 10 warm-up steps, 6400 examples per corpus and seed 0, on the attention and MLP projections. 
Each fine-tune's step-0 adapter reproduces the base model's activations bit for bit at all 17 evaluated layers. 
We fixed this gate, activations within $10^{-4}$ at every layer (registered row~2, Supplement~\ref{supp:registry}), before any LoRA result existed. 
Training examples whose prompt is verbatim in a frozen split are removed before training: 7598 for HH-harmless and 2243 for PKU-SafeRLHF. 
That check reads the prompt field alone, and no training record shows a remaining prompt overlap; Section~\ref{sec:res-lora} reports the evaluation items that the HH-harmless conversations held in their other user turns, and Supplement~\ref{supp:provenance} gives our other provenance checks. 

\textit{Phase~2} has nine instruct-side updates, fixed by a registered rule before any Phase~2 evaluation score existed. 
Four are releases: Mistral-7B-Instruct v0.1$\to$v0.2, Llama-3-8B-Instruct$\to$Llama-3.1-8B-Instruct, Qwen2-7B-Instruct$\to$Qwen2.5-7B-Instruct, and Phi-3-mini-128k-instruct$\to$Phi-3.5-mini-instruct, which also changes the RoPE scaling. 
Five are post-training steps: OLMo-2-1124-7B base$\to$SFT$\to$DPO$\to$Instruct and Llama-3.1-Tulu-3-8B SFT$\to$DPO$\to$final. 
Mistral-7B-Instruct v0.2$\to$v0.3 changes the tokeniser; we report it outside the headline set. 
Both that exclusion and the RoPE-scaling caveat were set before the first Phase~2 evaluation score. 
Phase~2 uses WildGuardMix alone, with 5000 calibration negatives and an evaluation split of 6000 negatives and 4000 positives. 
Aegis~2.0 is not used in Phase~2, because too few of its benign prompts lie outside Phase~1. 
In three Phase~2 pairs the rendered input changes with the weights, and we do not separate the two effects. 
Llama-3-8B-Instruct$\to$Llama-3.1-8B-Instruct adds a default system block, Qwen2-7B-Instruct$\to$Qwen2.5-7B-Instruct changes the default system text, and OLMo-2-1124-7B base$\to$SFT goes from raw text to the SFT chat template. 
Because the data, the pooling and the kind of update all differ from Phase~1, we treat Phase~2 as a second population, not an exact replication. 

\textit{Phase~3} runs its model-side attacks and its defences on one backbone, Mistral-7B-Instruct-v0.2; A1 uses the Phase~1 cells (Section~\ref{sec:attacks}). 
The clean update is a rank-16 LoRA trained on Alpaca for 400 steps, with three seeds. 

By extraction record, the pre- and post-update sides of 14 of the 18 headline cells ran on the same GPU type: a V100 in Phase~1 and an A100 in Phase~2; Section~\ref{sec:disc-lim} gives the other four. 
In the hardware control the A100 extraction's activations are bit-identical to the V100's at only 1 of the 17 evaluated layers, so the registered rule that allowed at most $10^{-4}$ per layer FAILED (registered row~13, Supplement~\ref{supp:registry}). 
Yet at all 17 layers the control's paired ratio stays between 0.9995 and 1.0029 with the selected $C$ and between 0.9994 and 1.0019 with $C=1$, and its bands within [0.9857, 1.0200] and [0.9846, 1.0222], against the cut at 2 and 0.5 that defines damage. 
We read the headline cells' hardware and the control's ranges after the results existed, so both are descriptive. 
The rule of row~13 decided whether our LoRA dose series stood as measured, since its first runs had read the base model and the 25-step adapter on a V100 and the 100- and 400-step adapters on an A100. 
As the rule required when it failed, we re-extracted the base model and the 25-step adapter on the A100 and reran the three dose analyses; no registered verdict changed, and every LoRA number we report has both sides extracted on an A100. 

\subsection{Metrics and damage rules}\label{sec:metrics}

\textit{Operating point.} When we call the frozen guardrail \emph{calibrated}, we mean its operating point: the false-positive and true-positive rates of the frozen threshold $T$. 
This is not probability calibration, the expected calibration error that Duan~\cite{duan2026monitors} also records, and it is not Duan's recall at 1\% and 5\% false-positive rate. 
For each probe seed we compute the false-positive rate on the same evaluation negatives before and after the update, each floored at $1/n_{\text{neg}}$, where $n_{\text{neg}}$ is the number of evaluation negatives, and take their ratio. 
The paired ratio $r$ is the mean of this ratio over seeds and then over draws. 
On a draw, a cell is \emph{damaged} if its paired ratio is at least 2 (false positives at least double: the guardrail fails closed) or at most 0.5 (it fails open). 
A cell's class (fail closed, fail open or within) applies the same cut once, to its mean paired ratio over the draws. 
We write FPR/target for the realised false-positive rate divided by its target, so a value of 1 is on target. 
We report the true-positive rate at $T$ before and after the update; its interval resamples both draws and items (5000 replicates), an interval on AUROC resamples items, and the bands of Section~\ref{sec:probe} are percentiles over re-partitions of fixed data with no item resampling. 

\textit{Ranking.} We report AUROC on the evaluation split and its change under the update, with a paired item bootstrap over positives and negatives ($B=5000$). 
AUROC and CKA use no threshold, so the values we report are read on the fixed evaluation split, not re-partitioned, except where the realignment repair and the attacks read AUROC on each draw. 
As the AUROC-based check a deployer might run, we take the three criteria of Duan~\cite{duan2026monitors} (arXiv:2606.15980v2, 18 June 2026). 
A cell is eligible if its pre-update AUROC is at least 0.75; this is an eligibility gate, not a staleness trigger. 
A cell has a big drop if its change in AUROC is at most $-0.05$, and fails operationally if its post-update AUROC is below 0.70. 
An eligible cell is \emph{stale} if it has a big drop or fails operationally; Duan applies the criteria to each probe seed, and we apply them to the mean over the five probe seeds. 
The AUC rule flags a cell with a big drop or an operational failure; it does not apply the eligibility gate. 

\textit{Over-flagging of safe prompts.} The probe false-flag rate on XSTest~\cite{rottger2024xstest} safe prompts is the share of XSTest's 250 safe prompts that the frozen probe flags at $T$. 
A flag is the probe's decision, not a model refusal. 
We report its paired ratio as above, floored at $1/250$, and the 200 unsafe contrast prompts alongside. 

\textit{Category blinding.} For the targeted attack (Section~\ref{sec:attacks}), a cell is \emph{blinded} when the target category's true-positive rate on the evaluation split falls by at least 10 points at both the frozen $T$ and the threshold re-set on the stored calibration set, $R_{\text{same}}$ (Section~\ref{sec:repairs}). 
A blinded cell is \emph{stealthy} if, in addition, the attacked model stays inside the stealth bound, the AUC rule stays silent, and the estimated-rate check (Section~\ref{sec:twomoment}) stays silent on the calibration set. 
A category is a \emph{successful category} if at least 3 of its 5 attack seeds give a stealthy cell at the stronger shift, $c=-2$ (Section~\ref{sec:attacks}). 

\textit{Registration.} Each result is labelled as registered, with its verdict, or as descriptive; Supplement~\ref{supp:registry} lists every registered prediction and its verdict, and names of the form \mbox{P-\textit{x}.\textit{n}} and \mbox{P2-\textit{x}.\textit{n}} are the registry's names for its entries.
\emph{Registered} means committed as a dated entry to our own repository, not filed with a public registry; where that commit does not precede the result, we say so.

\subsection{The two-moment model and the estimated-rate check}\label{sec:twomoment}

Both tools describe the benign scores by their mean and standard deviation, under the same frozen probe and threshold.
On draw $d$ the pooled negatives split into a calibration part $C_d$ and an evaluation part $E_d$, and $T_{s,d}$ is the pre-update model's 1\% threshold on $C_d$ for probe seed $s$. 
We write $f_{\text{pre}}$ and $f_{\text{post}}$ for the frozen probe's scores on the pre- and post-update model. 
Positives enter neither tool. 
Every false-positive rate below is taken on $E_d$ at $T_{s,d}$ and floored at $1/|E_d|$. 

\textit{The two-moment model.} It asks whether the change in the negatives' mean and standard deviation accounts for the change in the false-positive rate.
For each seed it takes $\mu_0,\sigma_0$ and $\mu_1,\sigma_1$, the mean and standard deviation of $f_{\text{pre}}$ and of $f_{\text{post}}$ over the evaluation negatives $E_d$, and maps every pre-update score $x\in E_d$ to 
\begin{equation}
  \tilde f(x)=\mu_1+\bigl(f_{\text{pre}}(x)-\mu_0\bigr)\,\frac{\sigma_1}{\sigma_0}.
  \label{eq:twomoment}
\end{equation}
Its modelled paired ratio $\hat r_d$ is the realised one, $r_d$, with $\tilde f$ in place of $f_{\text{post}}$: 
\begin{equation}
  \hat r_d=\frac{1}{5}\sum_{s}\frac{\mathrm{FPR}(\tilde f)}{\mathrm{FPR}(f_{\text{pre}})},\qquad
  r_d=\frac{1}{5}\sum_{s}\frac{\mathrm{FPR}(f_{\text{post}})}{\mathrm{FPR}(f_{\text{pre}})}.
  \label{eq:ratios}
\end{equation}
Per cell, $\hat R$ and $R$ are the means of $\hat r_d$ and $r_d$ over the draws on which both are finite, which is every draw, with the fixed split excluded. 
The error is $e=|\log R-\log\hat R|$, in natural logarithms; a cell is within $\times1.5$ if $e\le\log1.5$, and a non-finite $e$ counts as a miss. 
We report the share of cells within $\times1.5$ alongside the median $e$. 
A translation-only comparator keeps $\sigma_1/\sigma_0=1$. 
The two-moment model uses the post-update moments of the same evaluation negatives whose false-positive rate it gives. 
It therefore explains the change; it does not forecast it.

\textit{The estimated-rate check.} It is the label-free check a deployer can run after an update, with the stored calibration negatives and the updated model. 
For each seed it takes $\mu^{C}_0,\sigma^{C}_0$ and $\mu^{C}_1,\sigma^{C}_1$, the mean and standard deviation of $f_{\text{pre}}$ and of $f_{\text{post}}$ over the calibration negatives $C_d$ only, and compares Gaussian tails at the threshold: 
\begin{equation}
  G_d=\frac{1}{5}\sum_{s}\frac{\bar\Phi\bigl((T_{s,d}-\mu^{C}_{1})/\sigma^{C}_{1}\bigr)}{\bar\Phi\bigl((T_{s,d}-\mu^{C}_{0})/\sigma^{C}_{0}\bigr)},\qquad
  \bar\Phi=1-\Phi.
  \label{eq:fourc}
\end{equation}
The flag fires when $G_d\ge2$ or $G_d\le0.5$, the cut of the damaged class, and it is scored against the cell's damaged class on the same draw. 
$C_d$ and $E_d$ are disjoint in every draw, so the check is out of sample: no evaluation negative behind the realised ratio enters the moments behind the flag. 
We report it descriptively. 
The AUC rule is scored against the same class; it uses no threshold, so its flag is the same on every draw. 

\subsection{Repairs}\label{sec:repairs}

Repairs use no new labels.
The threshold re-set recomputes the 1\% threshold on the post-update scores of 1000 stored calibration negatives ($R_{\text{same}}$) or of 1000 negatives held out of the draw's evaluation part ($R_{\text{fresh}}$), exchangeable with the rest of that part. 
Quantile transfer carries $T$'s rank among the pre-update scores of 1000 unlabelled traffic prompts, 1\% of them harmful, to the same rank among their post-update scores. 
Restandardisation re-estimates each feature's mean and standard deviation on post-update prompts, class-balanced or traffic, and keeps $w$, $b$ and $T$.
Realignment maps post-update states back into the pre-update space with a ridge or an orthogonal Procrustes map~\cite{schonemann1966procrustes}, since Duan~\cite{duan2026monitors} writes a map $A$ fitted on paired activations but not the fitting method. 
The three maps of Section~\ref{sec:res-mech} that test the mechanism, the paired alignment map, Procrustes and CORAL, are fitted from post- to pre-update states on the probe's training prompts and the 1000 harmful prompts that the fixed calibration split holds beside its 1000 negatives, never on those negatives. 
Supplement~\ref{supp:details} gives the z-score transfer, the map fits and the labelled refit, which is an upper bound only.

\subsection{Attacks and the canary}\label{sec:attacks}

Attack sizes $c$ are in units of $\sigma$, the standard deviation of a probe seed's pre-update calibration-negative scores, taken per draw in A1 and on the fixed split in A1m and A2, which are trained once. 

\textit{A1} adds $\delta=c\,\sigma\,v/(v^\top v)$, with $v=w/\mathrm{sd}$ and $\mathrm{sd}$ the probe's standardisation vector of Section~\ref{sec:probe}, to every post-update state of the nine Phase~1 cells at $L^*$, so every frozen score moves by $c\sigma$ and AUROC does not change. 
For A1's misses, let $\kappa$ be the amount by which the post-update evaluation negatives' 99th percentile exceeds a Gaussian's, in standard units. 
\textit{A1m} produces this translation with a rank-1 LoRA trained on top of the clean update. 
\textit{A2} uses A1m's recipe but shifts only one target category's training prompts by $c\sigma$; every other prompt has a zero-shift term, and a KL term keeps the model near the clean update. 
The targets are cat-1 to cat-3, the three categories with the most items, each with $c\in\{-2,-1\}$ and five seeds: 30 attacked models. 
A model is inside the \emph{stealth bound} if its KL from the clean update on held-out benign prompts is at most the clean seeds' seed-to-seed KL (0.08644 nats per token) and its ARC-Easy accuracy is within the clean seeds' range. 
\textit{A3} poisons a share $\rho$, from 0.001 to 0.25, of a repair's 1000-item reference. 
We run it on the nine Phase~1 cells against four combinations of reference and repair: $R_{\text{same}}$ and $R_{\text{fresh}}$ with the threshold re-set, and traffic at 1\% prevalence with quantile transfer and with restandardisation. 
A3 thus has 36 cells, one for each Phase~1 cell and combination. 
The poison is high-scoring harmful or benign prompts, or harmful prompts chosen to keep the estimated-rate check inside $(0.5,2)$. 

\textit{The canary.} For each target category the deployer holds 47 to 99 labelled harmful prompts that no attack or evaluation split uses. 
Per draw, the canary statistic is the largest of the three categories' falls in canary true-positive rate at the frozen threshold, each fall averaged over the five probe seeds.
It fires at $\tau$ or above, with $\tau$ set to match the global estimated-rate check's clean false-alarm count: the smallest value that at most that many of the 603 clean statistics reach. 
We fixed the canary after A2 was scored, so it is a post hoc check, and no attack adapted to it. 
Supplement~\ref{supp:details} gives each attack's grid and training settings.
\space%
\begin{table}[t]
\caption{In the headline cells the frozen threshold's operating point moves in both directions while AUROC barely changes.}
\label{tab:headline}
\centering\scriptsize\setlength{\tabcolsep}{0pt}
\begin{tabular*}{\columnwidth}{@{\extracolsep{\fill}}lllrlrlrrrr@{}}
\toprule
 & & & \multicolumn{2}{c}{1\% target} & \multicolumn{2}{c}{5\% target} & \multicolumn{2}{c}{TPR at 1\%} & \multicolumn{2}{c}{AUROC} \\
\cmidrule(lr){4-5}\cmidrule(lr){6-7}\cmidrule(lr){8-9}\cmidrule(l){10-11}
ID & Data & $L^*$ & $r$ & class & $r$ & class & pre & post & pre & post \\
\midrule
1a & Aegis & 12 & 17.377 & closed & 5.168 & closed & 0.396 & 0.791 & 0.9170 & 0.9112 \\
 &  &  & \multicolumn{2}{l}{[10.20, 32.88]} & \multicolumn{2}{l}{[4.19, 6.40]} &  &  &  &  \\
1a & WGM & 14 & 9.521 & closed & 3.476 & closed & 0.713 & 0.894 & 0.9704 & 0.9665 \\
 &  &  & \multicolumn{2}{l}{[6.03, 16.58]} & \multicolumn{2}{l}{[2.88, 4.29]} &  &  &  &  \\
1b & Aegis & 18 & 6.783 & closed & 2.955 & closed & 0.413 & 0.681 & 0.9199 & 0.9213 \\
 &  &  & \multicolumn{2}{l}{[4.44, 11.49]} & \multicolumn{2}{l}{[2.47, 3.54]} &  &  &  &  \\
1c & Aegis & 14 & 2.395 & closed & 1.779 & within & 0.420 & 0.541 & 0.9163 & 0.9211 \\
 &  &  & \multicolumn{2}{l}{[1.96, 3.13]} & \multicolumn{2}{l}{[1.61, 1.98]} &  &  &  &  \\
1c & WGM & 14 & 2.081 & closed & 1.559 & within & 0.696 & 0.798 & 0.9691 & 0.9741 \\
 &  &  & \multicolumn{2}{l}{[1.70, 2.63]} & \multicolumn{2}{l}{[1.40, 1.79]} &  &  &  &  \\
1d & Aegis & 14 & 2.152 & closed & 1.834 & within & 0.411 & 0.535 & 0.9170 & 0.9212 \\
 &  &  & \multicolumn{2}{l}{[1.83, 2.65]} & \multicolumn{2}{l}{[1.66, 2.06]} &  &  &  &  \\
1e & Aegis & 18 & 1.577 & within & 1.225 & within & 0.371 & 0.427 & 0.9093 & 0.9133 \\
 &  &  & \multicolumn{2}{l}{[1.34, 1.95]} & \multicolumn{2}{l}{[1.17, 1.29]} &  &  &  &  \\
1f & Aegis & 14 & 3.146 & closed & 2.105 & closed & 0.380 & 0.570 & 0.9168 & 0.9211 \\
 &  &  & \multicolumn{2}{l}{[2.39, 4.75]} & \multicolumn{2}{l}{[1.86, 2.40]} &  &  &  &  \\
1g & Aegis & 14 & 3.224 & closed & 1.803 & within & 0.420 & 0.570 & 0.9163 & 0.9161 \\
 &  &  & \multicolumn{2}{l}{[2.46, 4.71]} & \multicolumn{2}{l}{[1.60, 2.03]} &  &  &  &  \\
\addlinespace
2a & WGM & 16 & 0.055 & open & 0.109 & open & 0.738 & 0.365 & 0.9776 & 0.9733 \\
 &  &  & \multicolumn{2}{l}{[0.03, 0.09]} & \multicolumn{2}{l}{[0.09, 0.13]} &  &  &  &  \\
2b & WGM & 14 & 2.254 & closed & 1.642 & within & 0.773 & 0.839 & 0.9814 & 0.9808 \\
 &  &  & \multicolumn{2}{l}{[1.98, 2.62]} & \multicolumn{2}{l}{[1.56, 1.74]} &  &  &  &  \\
2c & WGM & 18 & 1.028 & within & 0.699 & within & 0.743 & 0.697 & 0.9787 & 0.9747 \\
 &  &  & \multicolumn{2}{l}{[0.90, 1.19]} & \multicolumn{2}{l}{[0.65, 0.75]} &  &  &  &  \\
2d & WGM & 18 & 1.171 & within & 1.175 & within & 0.732 & 0.741 & 0.9777 & 0.9767 \\
 &  &  & \multicolumn{2}{l}{[1.09, 1.27]} & \multicolumn{2}{l}{[1.14, 1.21]} &  &  &  &  \\
2e & WGM & 14 & 5.874 & closed & 3.703 & closed & 0.710 & 0.898 & 0.9730 & 0.9760 \\
 &  &  & \multicolumn{2}{l}{[4.92, 7.12]} & \multicolumn{2}{l}{[3.38, 4.08]} &  &  &  &  \\
2f & WGM & 16 & 1.050 & within & 1.009 & within & 0.829 & 0.837 & 0.9860 & 0.9864 \\
 &  &  & \multicolumn{2}{l}{[1.00, 1.10]} & \multicolumn{2}{l}{[0.99, 1.03]} &  &  &  &  \\
2g & WGM & 16 & 1.057 & within & 1.046 & within & 0.831 & 0.835 & 0.9863 & 0.9863 \\
 &  &  & \multicolumn{2}{l}{[1.03, 1.09]} & \multicolumn{2}{l}{[1.04, 1.06]} &  &  &  &  \\
2h & WGM & 14 & 0.779 & within & 0.628 & within & 0.825 & 0.778 & 0.9866 & 0.9859 \\
 &  &  & \multicolumn{2}{l}{[0.71, 0.86]} & \multicolumn{2}{l}{[0.60, 0.66]} &  &  &  &  \\
2i & WGM & 14 & 0.905 & within & 0.896 & within & 0.828 & 0.814 & 0.9868 & 0.9867 \\
 &  &  & \multicolumn{2}{l}{[0.87, 0.94]} & \multicolumn{2}{l}{[0.88, 0.91]} &  &  &  &  \\
\bottomrule
\end{tabular*}
\par\smallskip\parbox{\columnwidth}{\scriptsize $r$: the paired ratio FPR$_{\mathrm{post}}$/FPR$_{\mathrm{pre}}$ at the frozen threshold, mean over the 1000 re-partitions, each with 2000 evaluation negatives in Phase~1 (1a to 1g) and 6000 in Phase~2 (2a to 2i); beneath each row, the band of each $r$, its 2.5th and 97.5th percentiles over the re-partitions; class: closed if $r\ge2$, open if $r\le0.5$, within otherwise; TPR at the frozen 1\% threshold, mean over the re-partitions; AUROC before the update and of the frozen probe after it, on the fixed split, mean over the five probe seeds. Aegis: Aegis~2.0; WGM: WildGuardMix.}
\end{table}

\begin{table}[t]
\caption{At matched dose the two safety-trained adapters move the frozen threshold more than the general one, and their shifts are both longer than the general arm's and have a larger $|\cos\theta|$ with the probe direction, though every shift is nearly orthogonal to it.}
\label{tab:lora}
\centering\scriptsize\setlength{\tabcolsep}{0pt}
\begin{tabular*}{\columnwidth}{@{\extracolsep{\fill}}lrrrrrrr@{}}
\toprule
 & & & \multicolumn{2}{c}{worst FPR/target} & \multicolumn{3}{c}{at $L^*=12$} \\
\cmidrule(lr){4-5}\cmidrule(l){6-8}
Corpus & Steps & Frozen error & ratio & layer & $\Delta$TPR & $\cos\theta$ & $\lVert d\rVert$ \\
\midrule
Alpaca & 25 & 0.446 & 1.60 & 10 & $+$0.039 & $-$0.0013 & 28.88 \\
 &  & [0.251, 0.850] & \multicolumn{2}{l}{[0.70, 2.64]} &  &  &  \\
 & 100 & 0.311 & 1.44 & 14 & $+$0.034 & $-$0.0014 & 17.49 \\
 &  & [0.158, 0.630] & \multicolumn{2}{l}{[0.64, 2.36]} &  &  &  \\
 & 400 & 0.381 & 1.58 & 14 & $+$0.029 & $-$0.0030 & 16.14 \\
 &  & [0.202, 0.770] & \multicolumn{2}{l}{[0.74, 2.57]} &  &  &  \\
\addlinespace
HH-harmless$^{\S}$ & 25 & 2.613 & 34.34 & 14 & $+$0.375 & $+$0.0592 & 59.02 \\
 &  & [2.455, 2.764] & \multicolumn{2}{l}{[29.39, 39.40]} &  &  &  \\
 & 100 & 1.901 & 12.30 & 24 & $+$0.220 & $+$0.0486 & 27.18 \\
 &  & [1.667, 2.103] & \multicolumn{2}{l}{[8.86, 15.20]} &  &  &  \\
 & 400 & 1.815 & 12.08 & 30 & $+$0.213 & $+$0.0565 & 23.31 \\
 &  & [1.575, 2.016] & \multicolumn{2}{l}{[9.80, 14.73]} &  &  &  \\
\addlinespace
PKU-SafeRLHF & 25 & 2.340 & 21.33 & 18 & $+$0.274 & $+$0.0408 & 51.18 \\
 &  & [2.112, 2.538] & \multicolumn{2}{l}{[16.70, 25.98]} &  &  &  \\
 & 100 & 1.529 & 10.49 & 32 & $+$0.105 & $+$0.0352 & 22.27 \\
 &  & [1.235, 1.778] & \multicolumn{2}{l}{[8.39, 12.69]} &  &  &  \\
 & 400 & 1.845 & 12.27 & 30 & $+$0.190 & $+$0.0460 & 25.11 \\
 &  & [1.585, 2.056] & \multicolumn{2}{l}{[9.80, 14.94]} &  &  &  \\
\bottomrule
\end{tabular*}
\par\smallskip\parbox{\columnwidth}{\scriptsize LoRA fine-tunes of Mistral-7B v0.1 on Aegis~2.0. Frozen error: mean $|\log(\mathrm{FPR}/\mathrm{target})|$ over the layer sweep; worst FPR/target: the largest per-layer mean over the sweep and its layer; beneath each run, the bands of both, their 2.5th and 97.5th percentiles over the re-partitions; $\Delta$TPR: post minus pre at the frozen threshold; $\cos\theta$ and $\lVert d\rVert$: the signed cosine between the mean shift of the fixed evaluation split's negatives and the probe direction, and the shift's length in standard deviations, $d=(\mu_{\mathrm{post}}-\mu_{\mathrm{pre}})/\mathrm{sd}_{\mathrm{pre}}$: the negatives' mean hidden state after minus before the update, divided feature by feature by the standard deviation of the probe's pre-update fit prompts; all means over the re-partitions or the probe seeds. For a random direction in 4,096 dimensions, $|\cos\theta|$ has mean 0.01239 (analytically $\sqrt{2/(4096\pi)}=0.01247$) and 95\% band [0.00049, 0.03489], pooled over 10,000 directions for each probe seed; each seed's observed $|\cos\theta|$ lies at percentiles 1.2 to 43.7 of its own random-direction distribution for Alpaca and 93.3 to 100.0 for the safety arms (HH-harmless 97.4 to 100.0, PKU-SafeRLHF 93.3 to 100.0). $^{\S}$ Trained on conversations that contain evaluation items (see text).}
\end{table}

\begin{table}[t]
\caption{Per Phase~1 cell the estimated-rate check flags most damaged layer cells and few undamaged ones, except in 1e, where it misses most of the few damaged cells (3.23 per draw), and in 1g, where it flags most of the few undamaged cells (3.00 per draw).}
\label{tab:fourc}
\centering\scriptsize\setlength{\tabcolsep}{0pt}
\begin{tabular*}{\columnwidth}{@{\extracolsep{\fill}}llrrrr@{}}
\toprule
 & & \multicolumn{2}{c}{damaged (per draw)} & \multicolumn{2}{c}{undamaged (per draw)} \\
\cmidrule(lr){3-4}\cmidrule(l){5-6}
ID & Data & layer cells & flagged & layer cells & flagged \\
\midrule
1a & Aegis & 15.98 & 15.98 & 1.02 & 0.02 \\
1a & WGM & 16.00 & 16.00 & 1.00 & 0.00 \\
1b & Aegis & 17.09 & 17.00 & 4.91 & 0.00 \\
1c & Aegis & 10.85 & 9.57 & 6.15 & 0.31 \\
1c & WGM & 6.50 & 4.83 & 10.50 & 2.01 \\
1d & Aegis & 12.30 & 9.65 & 4.70 & 0.93 \\
1e & Aegis & 3.23 & 1.00 & 11.77 & 0.00 \\
1f & Aegis & 7.08 & 6.53 & 9.92 & 0.10 \\
1g & Aegis & 14.00 & 14.00 & 3.00 & 2.00 \\
\bottomrule
\end{tabular*}
\par\smallskip\parbox{\columnwidth}{\scriptsize Means over the 1000 re-partitions of the number of the cell's evaluated layers in each class on a draw. Aegis: Aegis~2.0; WGM: WildGuardMix.}
\end{table}
\space     %
\begin{figure}[t]
\centering
\begin{tikzpicture}
\begin{axis}[width=0.97\columnwidth, height=0.72\columnwidth, xmode=log, ymode=log,
  log ticks with fixed point, xtick={0.001,0.01,0.1,0.5,2,10}, ytick={0.1,0.5,1,2,10,100},
  extra x ticks={0.5,2}, extra y ticks={0.5,2}, extra x tick labels={}, extra y tick labels={},
  extra tick style={grid=major, grid style={black!45, line width=0.5pt}},
  xlabel={Estimated rate ratio $G_d$}, ylabel={Realised paired ratio $r$},
  tick label style={font=\scriptsize}, label style={font=\footnotesize},
  legend style={font=\scriptsize, draw=none, fill=none}, legend pos=north west, legend cell align=left]
\addplot[only marks, mark=o, mark size=1.2pt, black] coordinates {
  (1.2287,1.286) 
  (28.2532,102.036) 
  (30.3889,94.500) 
  (17.0375,25.633) 
  (13.2021,21.557) 
  (26.9007,34.977) 
  (25.0432,38.980) 
  (22.3354,45.455) 
  (26.3508,47.222) 
  (17.2247,34.970) 
  (12.5849,24.056) 
  (9.3775,15.547) 
  (11.6646,12.817) 
  (10.3736,9.987) 
  (8.5836,9.147) 
  (6.6342,7.748) 
  (0.3678,0.401) 
  (1.2597,1.249) 
  (5.4597,19.538) 
  (6.3849,20.572) 
  (4.1795,9.452) 
  (3.9269,7.775) 
  (7.1056,12.398) 
  (7.3568,9.483) 
  (5.7152,8.194) 
  (10.0559,11.010) 
  (9.5872,15.764) 
  (7.8196,9.059) 
  (5.1766,7.907) 
  (6.1121,8.424) 
  (7.9620,9.325) 
  (7.4498,7.932) 
  (4.9479,5.995) 
  (0.1512,0.176) 
  (1.1070,1.236) 
  (0.9954,1.275) 
  (0.9832,1.138) 
  (1.4887,1.511) 
  (1.7186,2.978) 
  (2.8973,3.032) 
  (3.1725,3.668) 
  (6.9290,8.442) 
  (5.9670,13.033) 
  (8.0948,12.208) 
  (8.5916,11.677) 
  (9.8108,13.305) 
  (7.9470,12.344) 
  (5.8497,7.216) 
  (7.1873,6.498) 
  (5.4324,6.129) 
  (5.1014,5.408) 
  (4.2708,4.394) 
  (3.3396,3.646) 
  (3.3541,3.459) 
  (2.6695,2.681) 
  (0.0005,0.083) 
  (0.9822,1.050) 
  (1.2888,1.795) 
  (1.4007,2.382) 
  (1.3564,2.426) 
  (2.8983,3.562) 
  (3.1865,3.832) 
  (3.7376,4.101) 
  (2.4522,2.700) 
  (1.2757,1.244) 
  (0.7379,0.666) 
  (0.5581,0.503) 
  (0.4461,0.333) 
  (0.3254,0.232) 
  (0.2463,0.152) 
  (0.2186,0.154) 
  (0.2400,0.106) 
  (0.4731,0.303) 
  (1.0338,1.069) 
  (0.8410,0.534) 
  (0.6676,0.897) 
  (0.8202,1.141) 
  (0.8928,1.545) 
  (1.4289,2.491) 
  (3.0882,4.250) 
  (1.2890,2.085) 
  (0.8110,1.278) 
  (0.6719,1.189) 
  (0.5077,1.011) 
  (0.4196,0.954) 
  (0.2150,0.536) 
  (0.1543,0.480) 
  (0.0973,0.362) 
  (0.0194,0.183) 
  (0.1425,0.661) 
  (0.9838,0.858) 
  (1.2910,2.435) 
  (1.6541,3.250) 
  (1.4619,2.525) 
  (2.3591,3.427) 
  (2.4677,4.499) 
  (2.9724,4.414) 
  (2.0879,2.414) 
  (0.8235,0.976) 
  (0.5067,0.517) 
  (0.4108,0.455) 
  (0.2846,0.290) 
  (0.1871,0.196) 
  (0.1358,0.194) 
  (0.1012,0.175) 
  (0.1410,0.149) 
  (0.2745,0.178) 
  (1.0308,0.997) 
  (0.9159,0.802) 
  (0.9286,3.057) 
  (0.9938,3.903) 
  (0.8713,1.801) 
  (1.0749,1.315) 
  (0.9383,0.343) 
  (0.9868,0.873) 
  (1.1723,2.029) 
  (1.0257,1.810) 
  (0.6551,1.134) 
  (0.7588,0.809) 
  (0.8017,0.853) 
  (0.9037,0.894) 
  (3.8562,3.349) 
  (0.9938,0.982) 
  (1.3160,1.267) 
  (1.3554,1.609) 
  (1.9265,2.878) 
  (2.3042,3.950) 
  (2.6822,3.096) 
  (3.2001,6.007) 
  (3.9884,6.550) 
  (3.4644,4.005) 
  (3.4521,2.718) 
  (1.7873,1.839) 
  (1.0186,1.179) 
  (0.6766,0.941) 
  (0.7949,1.016) 
  (0.6862,1.090) 
  (0.5306,0.911) 
  (3.6702,4.562) 
  (2.8910,10.490) 
  (2.3107,1.394) 
  (2.5883,2.308) 
  (3.6146,5.118) 
  (3.6390,3.628) 
  (3.5859,4.439) 
  (4.3543,3.552) 
  (4.0176,3.899) 
  (5.0629,4.455) 
  (4.4736,5.053) 
  (3.8810,3.334) 
  (4.2000,3.375) 
  (4.2770,3.472) 
  (3.9672,3.210) 
  (4.0096,3.947) 
  (3.2031,3.545) 
  (1.2346,1.244) 
};
\addlegendentry{layer cell}
\addplot[only marks, mark=square, mark size=3pt, thick, black] coordinates {
  (28.2532,102.036) 
  (5.4597,19.538) 
};
\addlegendentry{flagged by the AUC rule}
\end{axis}
\end{tikzpicture}
\caption{Realised paired ratio against the estimated rate ratio $G_d$ in the 156 Phase~1 layer cells, on the fixed split. Grey lines: the cuts at 0.5 and 2 on both axes. Cells damaged and flagged / damaged and not flagged / undamaged and flagged / undamaged and not flagged: fixed split, 97~/~13~/~4~/~42; per draw (the mean over the 1000 re-partitions, to one decimal), 94.6~/~8.5~/~5.4~/~47.6. Squares: the two cells the AUC rule flags, on the fixed split.}
\label{fig:fourc-scatter}
\end{figure}
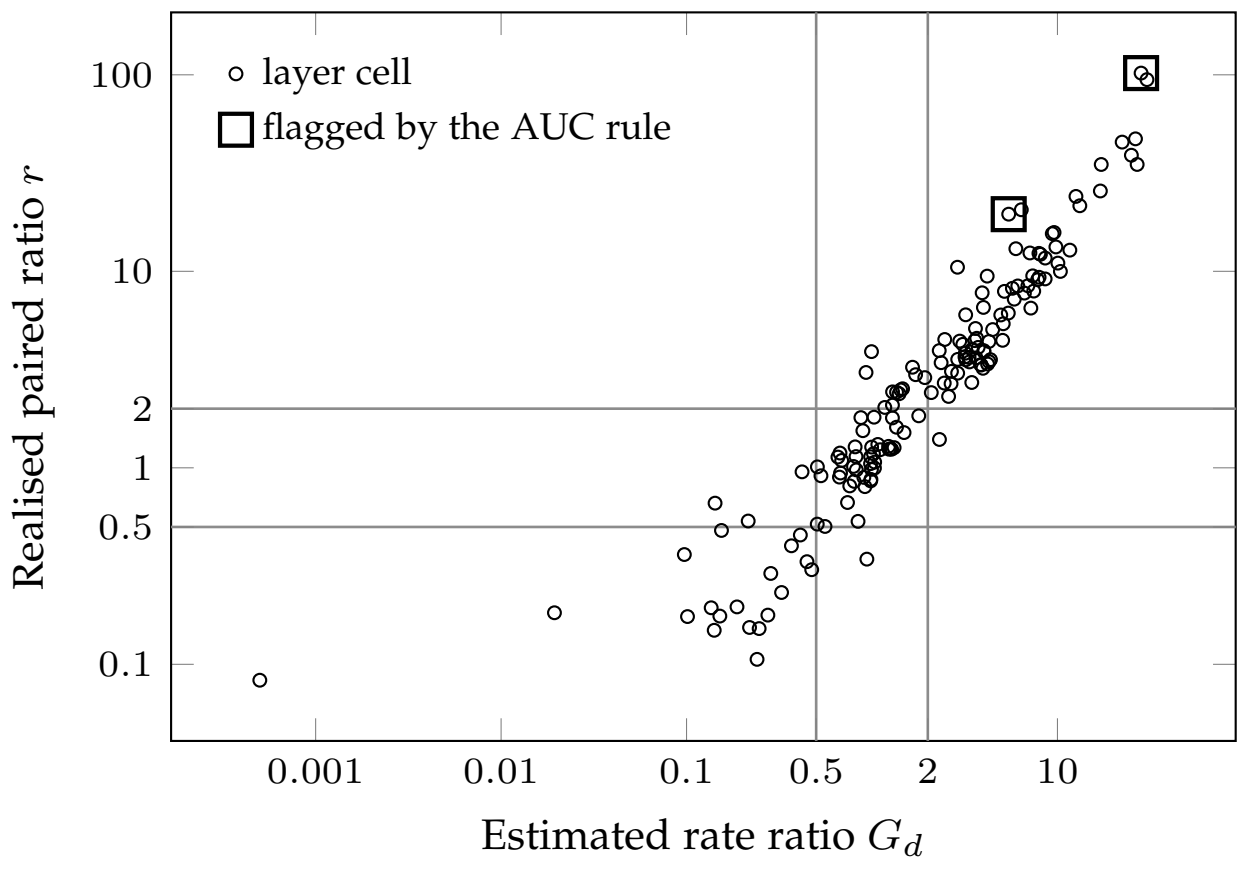
\space    %
\section{Results}\label{sec:results}

\subsection{Operating point moves, ranking does not}\label{sec:res-op}

With 5 probe seeds and 1000 re-partitions per cell (Section~\ref{sec:probe}), at 1\% the frozen threshold fails closed in 8 of the 9 Phase~1 headline cells and in 2 of the 9 Phase~2 cells, 2b and 2e, and fails open in one, 2a; the other 7 of the 18 cells are within (Table~\ref{tab:headline}). 
These classes are descriptive: we first wrote this classification of the headline cells after every Phase~1 and Phase~2 score existed, and no registered entry states it. 
We fixed the cuts at 2 and 0.5 after the first Phase~1 scores existed. 
At 1\% the bands of 1c on both datasets, 1d and 2b contain 2, and at 5\% those of 1d, 1f and 1g. 
We give the two phases apart because in our design the phase, the kind of update and the read-out coincide: every Phase~1 pair starts from a base model read on the raw prompt, six of the seven as base-to-instruct releases, while the Phase~2 pairs are instruct-side releases and post-training steps whose instruct models read their chat templates, so the difference between the phases cannot be credited to the kind of update.
On Aegis~2.0, the Mistral-7B v0.1 base-to-instruct update (1a) takes the frozen threshold's false-positive rate from 1.01\% before the update, on its 1\% target, to 15.12\% after it, while the probe's AUROC falls by 0.0058. 
Both rates are means over the re-partitions, and 1a's paired ratio in Table~\ref{tab:headline} averages the per-draw ratios, so it is not the ratio of these two means.
The band of 1a's paired ratio, [10.20, 32.88], lies wholly above the two-fold cut. 
In 2a, Mistral-7B-Instruct v0.1 to v0.2 on WildGuardMix, the band of the paired ratio, [0.025, 0.086], lies wholly below the cut at 0.5. 

Ranking does not move with it. 
No headline cell's AUROC changes by more than 1a's Aegis cell does, in either direction, and Duan's criteria call none of the 18 cells stale: every cell is eligible, and none has a big drop or fails operationally (Table~\ref{tab:headline}). 
On the nine LoRA runs at $L^*$ they call none stale on the seed mean; the worst change over the realignment study's draws, $-0.0470$, is in the HH-harmless run at 25 steps, an arm whose training data overlap the evaluation items, and on the fixed split one of that run's five probe seeds falls by 0.0550, a big drop (Section~\ref{sec:res-lora}). 
Over the 156 Phase~1 layer cells, the AUC rule, read on the fixed-split AUROC, flags two cells, 0.019 of those damaged on the average draw, at specificity 1.000. 

The move is not peculiar to one layer or one dataset. 
In the nine cells of 1a to 1g the maximum of the paired ratio over the layer sweep runs from 3.27 to 53.92 (Fig.~\mbox{\ref{fig:layer-sweep}}). 
For the extra pair 1h, whose paired ratio at $L^*$ is 1.01 [0.89, 1.19], the maximum is 2.89, so a maximum over layers crosses the two-fold cut even there. 
The Mistral update also fails closed on WildGuardMix, with a band of [6.03, 16.58] at its own $L^*$. 

In all 10 fail-closed cells the true-positive rate rises, so the failure is added false-positive load, not lost detection. 
One cell, 2a (configuration changes: Table~\ref{tab:config}), goes the other way: detection falls from 0.738 to 0.365, means over the re-partitions, while on the fixed split its AUROC moves from 0.9776 to 0.9733, and its rendered input does not change, so format does not confound it. 
Neither is visible from AUROC, so a deployer cannot tell them apart without re-measuring at the operating point. 

Representation similarity is not a reliable guide to the damage either.
In the ten cells of 1a to 1h the minimum linear centred kernel alignment (CKA) over layers never falls below 0.855, and it is 0.918 for 1a's Aegis cell. 
On Llama-3-8B to Llama-3.1-8B (1g) the minimum CKA reads 0.99302, yet the band of the paired ratio at $L^*=14$ is [2.46, 4.71]. 

Away from $L^*$ AUROC falls further, by at most 0.1057 over all layers (0.1565 with $C=1$, the sensitivity arm), and in eight of the ten cells of 1a to 1h its largest drop is below 0.05. 
The AUROC drop and the damage to the operating point, measured by the paired ratio over the re-partitions, show no firm cell-level link. 
Across the ten cells of 1a to 1h the Spearman correlation between the largest fixed-split AUROC drop and the largest paired ratio over layers is 0.430. 
On the five of these cells whose input caches were all verified by extraction record rather than by a pinned hash alone, 1b, 1d, 1f and the WildGuardMix cells of 1a and 1c, it is 0.400; we computed no interval for either. 
Across the 156 layer cells of 1a to 1g it is $+0.373$ [$-0.167$, $+0.669$], an interval from a bootstrap that resamples the nine cells of 1a to 1g, each with all its layer cells, and that includes 0. 

\subsection{Fragility depends on the target rate}\label{sec:res-five}

At 5\% the frozen threshold fails closed in 5 of the 18 headline cells and fails open in 2a, and 12 cells are within (descriptive: no registered prediction scores the 5\% arm; Table~\ref{tab:headline}). 
Five cells move from fail closed to within: 1c on both datasets, 1d, 1g and 2b. 
All five sit between 2.081 and 3.224 at 1\% and between 1.559 and 1.834 at 5\%, so they are cells crossing the cut at 2 between the two targets, not large effects vanishing. 

The paired ratio asks how far the false-positive rate moves; a second test asks whether the frozen threshold misses its target by more than the pre-update model itself does.
On each draw the \emph{pre-update dispersion} is the pre-update model's $|\log(\mathrm{FPR}_{\mathrm{pre}}/\mathrm{target})|$ at the draw's threshold, and the \emph{pre-update dispersion test} takes an arm's $|\log(\mathrm{FPR}/\mathrm{target})|$ at $L^*$ minus it. 
It detects damage when the 95\% band of that difference over draws lies wholly above 0, and it finds an arm within the pre-update dispersion when the band includes 0. 
At 1\% it detects damage in 2 of the 7 Aegis pairs, Mistral-7B v0.1 at L12 with band [1.662, 2.714] and Gemma-2 at L18 with band [0.739, 1.818]; the other 5 are within the pre-update dispersion. 
At 5\% it detects damage in 6 of the 7, all but Qwen2.5. 
At 5\% the pre-update dispersion also narrows: its standard deviation over draws is 0.36 to 0.45 of its 1\% value in each of the 7 pairs. 
The narrowing does not account for the shift: for Llama-3.1 the frozen arm's mean excess over the pre-update dispersion rises from 0.411 at 1\% to 0.462 at 5\%, while its paired ratio falls (Table~\ref{tab:headline}). 
We do not know why the two tests move apart at 5\%. 

\subsection{Mechanism: two moments and the estimated-rate check}\label{sec:res-mech}

The update does not destroy separability.
With $C=1$, refitting the probe on the post-update model changes its AUROC by a median within $\pm0.01$ while the frozen threshold's false-positive rate is off its target, which the rule we registered for this change reads as calibration-only damage. 
Taking each layer sweep as the median over its own layers, the change has a median of $+0.0020$ over 16 sweeps, the ten of 1a to 1h and six of the LoRA runs, 4 of them negative, against $-0.00002$ on the hardware control. 
The eight base-to-instruct sweeps among them have a median of $+0.0059$, all eight positive (two-sided exact sign test over the six pairs they come from, $p=0.031$; descriptive). 
Registered row~17 states this rule, but we committed it after the result it scores existed, so that result is descriptive. 
Duan~\cite{duan2026monitors} reports a related result: a probe retrained on the updated activations recovers a median 0.981 of the AUROC the frozen probe lost, over 42 cells. 
Concurrent work also finds probes refit on later checkpoints more separable than those of the base model~\cite{anon2026basecase,tsatsos2026probes}.

The two-moment model accounts for the change.
In Phase~1 the fit is in sample. 
Over 156 layer cells the slope of the log realised on the log modelled paired ratio is 0.993 [0.908, 1.095], 0.885 of the cells are within $\times1.5$, and the median error is 0.147. 
On the 120 of these cells whose realised paired ratio lies outside $\times1.5$ of 1, the model is within $\times1.5$ in 103 (0.858), median error 0.148, against 32 (0.267), median 0.860, for translation alone; on the 36 cells that do not move it is within in 35. 
We chose that subset on the realised value after the scores existed, so it is descriptive.
In Phase~2 we registered these predictions before any Phase~2 evaluation score existed. 
Over the 63 cells of the nine pairs, $L^*$ and the six grid layers, the slope is 0.911 [0.806, 1.292] (P2-M.1, HELD; Supplement~\ref{supp:registry}), and 0.905 of the cells are within $\times1.5$ with median error 0.058 (P2-M.2, HELD). 
Translation alone reaches 0.730 (P2-M.3, HELD), so the model meets all three criteria registered for Phase~2. 
On the 22 of these cells that move, chosen the same way after the scores, the model is within $\times1.5$ in 17 (0.773), median error 0.185, against 9 (0.409) for translation; this is below the registered 80\% line, but that line belongs to all 63 cells. 

Re-estimating only the representation mean accounts for 0.763 [0.640, 0.914] of the improvement that restandardisation makes at $L^*$, an improvement of 0.922 [0.484, 1.426] in $|\log(\mathrm{FPR}/\mathrm{target})|$, pooled over the seven Aegis pairs. 
Per pair the share is undefined in 5 of the 7, where the improvement's band includes zero (UNDEFINED). 
Under compression, concurrent work re-centres a frozen probe's inputs on the new model's mean activations, with the probe's weights fixed, and removes most of the fall in true-positive rate at its fixed threshold on its headline layers~\cite{anon2026frozenharm}.

To test the mechanism we also mapped post-update states back onto pre-update ones and kept the probe and threshold: a ridge map fitted on the same prompts through both models (the paired alignment map), an orthogonal Procrustes map fitted the same way, and CORAL~\cite{sun2016coral} with Ledoit-Wolf shrinkage~\cite{ledoit2004wellconditioned}. 
We had registered the prediction that the paired alignment map would be worse than no repair in at least 4 of 7 pairs; at $L^*$ it is worse in 0 of 7 (FAILED). 
Three predictions for these maps were registered for an earlier run, with $C=1$ at five layers per pair, and scored on the fixed split: one FAILED, one HELD and one was NOT SUPPORTED. 
The one that HELD predicted that Procrustes would calibrate better than the paired alignment map in at least 5 of 7 pairs. 
That run's maps were fitted differently from the current ones, so what follows is a new reading, not a reproduction: on the current maps at $C=1$ and the same five layers, Procrustes calibrates better than the paired alignment map in 7 of 7 pairs on the fixed split, the reading the prediction registered, and in 6 of 7 on the means over the re-partitions, with Gemma-2 the exception. 
With the selected probe over every layer, Procrustes calibrates better in 2 of 7 pairs on the means over the re-partitions, against 6 of 7 at $C=1$ over the same layers, so the readings differ by the probe's regularisation; no contrast between the maps was registered, so these comparisons are descriptive. 

The estimated-rate check flags the failures that the AUC rule misses.
Over the 156 Phase~1 layer cells of Section~\ref{sec:res-op}, it flags 0.918 of the damaged cells and clears 0.898 of the undamaged ones (Table~\ref{tab:fourc}, Fig.~\ref{fig:fourc-scatter}). 
These are point estimates from the mean counts per draw, with no interval. 
Counting the stored calibration negatives at or above $T$ before and after the update, the direct ratio, flags 0.936 of the damaged cells and clears 0.886 of the undamaged ones, computed the same way, so measuring the rate on the stored prompts also misclassifies some cells (descriptive). 
Against two stronger label-free baselines, the estimated-rate check has the better flag but not the better ranking.
The baselines are a two-sample Kolmogorov--Smirnov (KS) test~\cite{rabanser2019failing} between the pre- and post-update probe scores of the calibration negatives and average thresholded confidence (ATC)~\cite{garg2022leveraging} at the operating point, scored like the estimated-rate check. 
No baseline has both pooled sensitivity and pooled specificity at or above $G_d$'s (P-C5.3, HELD): KS reaches a sensitivity of 0.966 but a specificity of 0.543 (P-C5.1, HELD), and ATC a specificity above $G_d$'s but a sensitivity of 0.479. 
ATC ranks the cells better, with AUROC 0.965 against $G_d$'s 0.952, so P-C5.2, which predicted it would not, FAILED. 

\subsection{Over-flagging of safe prompts}\label{sec:res-xstest}

In Phase~2, with the threshold set on WildGuardMix and the input format unchanged, the frozen probe flags XSTest's safe prompts 2.178 times as often after Phi-3-mini to Phi-3.5-mini (2d), band [1.867, 2.733], and 0.265 times as often after Mistral-7B-Instruct v0.1 to v0.2 (2a), band [0.223, 0.383], the direction of 2a's fail-open cell. 
The four later post-training steps of OLMo-2 and Tulu-3, 2f to 2i, stay between 0.999 and 1.142, with every band within [0.883, 1.300]. 
In the three cells whose rendered input changes with the weights, the ratio is 5.197 [3.909, 5.929] for 2b, 2.421 [2.017, 2.667] for 2c and 10.397 [9.793, 11.667] for 2e, where OLMo-2's base model reads raw text and its SFT successor a chat template; there we cannot separate the format from the weights. 
All nine are descriptive: we registered the estimator, not a prediction. 
Tsatsos~\cite{tsatsos2026probes} reports rises of this kind for base-model probes on distilled successors, at the probes' default threshold.

In Phase~1 we measure over-flagging on seven pairs from five base checkpoints and four model families; OLMo-2 is not among them. 
The OLMo-2 pair is 1f, and it is outside the seven because no Phase~1 XSTest run exists for it: we first read its models on Aegis~2.0 alone, and by the time we added XSTest for the pairs first read that way, its model was no longer in our model cache. 
At each pair's own $L^*$, with its Aegis probe, the paired ratio is 13.50 [11.38, 15.88] for 1a, 6.15 [4.96, 8.15] for 1b, 1.79 [1.61, 2.16] for 1c, 2.03 [1.81, 2.28] for 1d, 1.25 [1.15, 1.33] for 1e, 2.72 [2.07, 3.40] for 1g and 0.90 [0.79, 1.02] for 1h. 
Of the seven, 1a shows the largest rise at mid-depth: at L16 the paired ratio is 9.43 [7.52, 12.42], and the mean share of the 250 safe prompts flagged goes from 0.053 to 0.480; the ratio averages per-draw ratios, so it is not the ratio of these two shares. 
On 1g, Llama-3-8B to Llama-3.1-8B, a version increment that also changes the RoPE scaling and the maximum positions (Table~\ref{tab:config}), the frozen threshold flags more safe prompts, with a paired ratio of 2.72 [2.07, 3.40] at $L^*=14$. 
Its XSTest caches mix hardware, a V100 before the update and an A100 after it; the hardware control, run on Mistral-7B v0.1 and not on this pair, bounds the effect of that change on the false-positive paired ratio (Section~\ref{sec:pairs}). 
On 1h, Mistral-7B v0.1 to v0.2, the band lies above 1 at L2, 3.328 [2.609, 4.337], at L4, 1.353 [1.245, 1.450], at L6, 1.111 [1.040, 1.191], and at L18, 1.122 [1.030, 1.207], while at L16 it is 1.000 [0.922, 1.096]; at eight of its other evaluated layers the band lies wholly below 1. 
Hardware does not explain these: Mistral-7B v0.1's largest V100-to-A100 difference in the share flagged, over 17 layers, is 0.000136, below the registered bound of 0.001 (HELD). 

\subsection{Repairs, including realignment}\label{sec:res-repair}

Re-setting the threshold on the post-update scores of stored ($R_{\text{same}}$) or fresh ($R_{\text{fresh}}$) benign prompts returns the false-positive rate to its target: at $L^*$ in the nine Phase~1 cells the mean FPR/target over the re-partitions is 0.98 to 1.02 on $R_{\text{same}}$ and 0.96 to 1.02 on $R_{\text{fresh}}$ (descriptive). 
The re-set targets the false-positive rate alone: its true-positive rate is the updated model's at that rate, 0.351 to 0.721 on $R_{\text{same}}$ and 0.350 to 0.722 on $R_{\text{fresh}}$ (descriptive). 
In 1a it lies below the pre-update rate of Table~\ref{tab:headline}, a mean over other draws, at 0.351 against 0.396 on Aegis~2.0 and 0.626 against 0.713 on WildGuardMix on $R_{\text{same}}$; in the other cells it stays close to that rate or above it. 
Quantile transfer on 1\% traffic gives FPR/target 0.98 to 1.09 when the traffic's prevalence is the same before and after the update. 

Restandardisation restores it only with a class-matched reference.
On the probe's class-balanced training rows it gives FPR/target 1.01 to 1.71 at $L^*$ across the nine Phase~1 cells and is within the pre-update dispersion in 7 of 7 Aegis pairs, pooled $+0.016$ [$-0.030$, $+0.092$] above it. 
At $L^*$ a correction of the first two moments thus restores the false-positive rate to within calibration resolution, and we describe the failure as a moment shift. 
Averaged over layers restandardisation shows a small pooled excess, $+0.039$ [$+0.008$, $+0.087$], that no single pair shows. 
On 1\% live traffic, a pool of 1000 prompts at 1\% prevalence, restandardisation overshoots, with FPR/target 4.74 to 9.82 on the Aegis pairs and 21.87 and 24.36 on WildGuardMix, and a benign-only reference does the same; either is worse than no repair in 8 of 9 cells. 
Registered row~50 states both this overshoot, about five- to tenfold on mostly benign traffic, and the return to target above, but we wrote it after the results it scores existed, so both results are descriptive. 

Repairs that read live traffic are exposed to its prevalence.
If the post-update window holds 5\% harmful prompts where the pre-update window held 1\%, quantile transfer cuts FPR/target to 0.12 to 0.31 and the true-positive rate to 0.205 to 0.229 on the seven Aegis pairs and to 0.312 and 0.320 on WildGuardMix; at 25\% the true-positive rate is 0.037 to 0.050, and 0.066 and 0.061. 

Each repair below loses true-positive rate when its reference is poisoned, the re-set on stored prompts included.
The high-score sets are the harmful prompts that the frozen probe scores highest after the update. 
Injected into a 1000-item reference, they first cost the threshold re-set 10 points of true-positive rate at 0.5\% in 4 pairs and at 1\% in 5, on stored and fresh references alike; at 1\%, ten items, the re-set flags 0.006 to 0.017 of harmful prompts in all nine Phase~1 cells, the count at which a closed form we registered puts its threshold above every benign prompt (registered row~49, HELD). 
Quantile transfer on 1\% traffic first loses 10 points at 1\% in all nine cells, where it still flags 0.153 to 0.489 of harmful prompts, and flags 0.009 to 0.018 at 2\%; restandardisation on traffic first loses 10 points at 10\% on the Aegis pairs and at 25\% on WildGuardMix. 
These rates are descriptive, and we ran this poisoning at $n=1000$ only. 

Realignment keeps AUROC by a criterion we extended from Duan's in 27 of 27 cells in each of the six arms, the ridge and the orthogonal Procrustes map on Phase~1, Phase~2 and LoRA (P-RA.1, HELD). 
Duan's repair rule covers stale cells only, and no frozen cell is stale, so we registered that the realigned probe must not be stale by Duan's two criteria. 
Each arm has nine units: the nine Phase~1 cells, the nine Phase~2 pairs or the nine LoRA runs. 
Each unit is scored on three reference pools, which gives the 27 cells of an arm. 
Fitted on $R_{\text{same}}$ with the probe and threshold frozen, realignment leaves FPR/target outside $\times1.5$ of the target in all 9 units of each arm with the ridge map, and in 6, 4 and 3 of 9 with Procrustes in Phase~1, Phase~2 and LoRA. 
We predicted that at least 6 of each arm's 9 units would stay outside $\times1.5$; the ridge arms and Procrustes in Phase~1 held this, and Procrustes in Phase~2 and LoRA failed it because fewer units stay outside (P-RA.2, FAILED). 
No contrast between the two map forms was registered, so their comparison is descriptive. 
The two-moment model's error on the realigned rate is scored per unit and reference pool, 27 per arm (P-RA.3). 
With Procrustes the model is within $\times1.5$ in every cell, share 1.000 of the 27 in Phase~1, Phase~2 and LoRA, with median errors 0.070, 0.021 and 0.057 (P-RA.3, HELD in all three). 
With the ridge map it reaches 1.000 in LoRA, median 0.236 (P-RA.3, HELD), but 0.556 in Phase~1 and 0.630 in Phase~2, medians 0.285 and 0.261, below the 80\% we predicted (P-RA.3, FAILED in both). 
The LoRA arms include the three HH-harmless runs, whose exposure Section~\ref{sec:res-lora} states.

\subsection{Probe-family generality}\label{sec:res-mlp}

We replace the linear probe with a one-hidden-layer multilayer perceptron (MLP) probe on the same cached features at the linear probe's $L^*$. 
It answers whether the failure of the operating point and its repair are specific to a linear probe: for this MLP, on the nine Phase~1 cells, they are not. 
We ran it after Phase~1 closed, so it can support Phase~1's results but not reopen them. 
Its frozen paired ratio is at least 2 in 7 of the 9 Phase~1 cells, against 8 for the linear probe (P-MLP.1, HELD; Supplement~\ref{supp:registry}), and its paired ratios are within $\times1.5$ of the linear probe's in all 9, with median $|\log|$ difference 0.085. 
Its AUROC changes by at most 0.0079 in either direction in all 9 (P-MLP.3, HELD). 
The threshold re-set on $R_{\text{same}}$ brings its FPR/target into $(0.5, 2)$ in 9 of 9 cells, as for the linear probe (P-MLP.2, HELD). 
The estimated-rate check's pooled sensitivity over the nine $L^*$ cells, not over every layer cell as above, is 0.807 for the MLP probe, against 0.860 for the linear probe and the registered 0.8 (P-MLP.4, HELD). 

\subsection{Controlled LoRA updates}\label{sec:res-lora}

One of the two safety corpora is exposed to our evaluation data.
The HH-harmless conversations its adapter consumed held frozen evaluation items in a user turn other than the prompt we checked: 46 conversations with 55 distinct items at 25 steps, 188 with 211 at 100 and 810 with 763 at 400, the last counted across every frozen split. 
At 400 steps those items are 0.079 of the 3000 pooled Aegis negatives and 0.061 of the 2000 evaluation positives. 
A wider match, a user turn whose normalised text equals a pooled negative or contains it, is an upper bound on that count that very short items inflate; at 400 steps it reaches 0.100 of the pooled negatives for HH-harmless, against 0.006 for Alpaca and 0.008 for PKU-SafeRLHF. 
No current record bounds paraphrase or near-duplicate contamination. 
The HH rows survive, at every dose, the removal of one registered set wider than the 763 above: the 826 items of the three Aegis files the comparison reads that a conversation consumed by 400 steps matched in any turn, exactly, normalised or contained (HELD, Supplement~\ref{supp:registry}). 
By the first count neither Alpaca nor PKU-SafeRLHF trained on a frozen evaluation item at any dose, so the PKU-SafeRLHF against Alpaca contrast does not rest on the exposure, and we treat PKU-SafeRLHF as the clean safety arm and mark the HH rows in Table~\ref{tab:lora}. 

At matched dose the safety-trained adapters damage the frozen probe more than the general one (Table~\ref{tab:lora}). 
Per draw, the safety arms' frozen error is 5.39 to 7.00 times the general arm's: 6.45 [2.92, 10.62], 7.00 [2.82, 11.97] and 5.39 [2.08, 9.01] for HH-harmless and 5.79 [2.52, 9.60], 5.63 [2.10, 9.54] and 5.48 [2.07, 9.13] for PKU-SafeRLHF at 25, 100 and 400 steps. 
We computed these ratios from the stored per-draw errors after the scores existed, so they are descriptive; the registered prediction that PKU-SafeRLHF's damage exceeds the general arm's at all three doses held on the fixed split (HELD). 
Each arm is one training run, with seed 0, so these bands come from re-partitions of the evaluation negatives and carry no variation between training runs. 
The worst FPR/target over layers puts the general arm below both safety arms at every dose. 
Damage does not grow with dose: the frozen error and the worst FPR/target are largest at 25 steps in every arm, and our registered prediction that damage is non-decreasing over 25, 100 and 400 steps, scored on the fixed split, is NOT SUPPORTED (H-LoRA-1). 
At L12 the true-positive rate rises far more under the two safety adapters than under the general one (Table~\ref{tab:lora}). 

At L12 the signed cosine $\cos\theta$ of the angle $\theta$ between the negatives' mean shift and the probe direction is slightly negative for Alpaca and positive for both safety arms (Table~\ref{tab:lora}). 
We read this as the safety updates moving benign prompts towards the probe's harmful side; it is an interpretation, not a demonstrated cause.
Three alignment predictions were registered on the fixed split. 
That PKU-SafeRLHF's $|\cos\theta|$ exceeds the general arm's at every dose HELD, and so did the prediction that the alignment term outweighs the shift's size. 
The third, that the safety arm's $|\cos\theta|$ is larger at a comparable shift length, is NOT SUPPORTED: the safety shifts were also longer, and the registered rule did not cover that case. 

Over the nine runs' layer cells the AUC rule flags 0.043 of the damaged cells, while the estimated-rate check flags 0.959 at specificity 0.988, and 0.975 at 0.987 on the six safety runs; at $L^*$ the AUC rule calls none of the runs stale (Section~\ref{sec:res-op}). 
On Alpaca its sensitivity is 0.111, over a denominator of 1.70 damaged layer cells per draw. 
On the items left after the HH-exposed ones are removed, the specificity on the six safety runs falls to 0.784, and we do not know why: our registered explanation, that most of the drop comes from cells whose flag changed while their class did not, FAILED (P-T4.1). 
Taken alone, cells that cross into the undamaged class, and so enter the specificity denominator, account for 0.664 of the drop (descriptive). 

\subsection{Attacks and the canary}\label{sec:res-attack}

\textit{A1.}
A1 leaves AUROC unchanged, as it must, and the AUC rule's flag never moves (P-A1.1 and P-A1.2, HELD; Supplement~\ref{supp:registry}). 
The two-moment model predicts A1's frozen false-positive rate within $\times1.5$ in only 0.656 of the cells, short of the registered 80\%, with every miss at $c<0$ (P-A1.3, FAILED). 
The estimated-rate check detects A1 with sensitivity 0.951 but specificity 0.568, so its registered prediction also FAILED (P-A1.4). 

\textit{The heavy-tail account.}
A registered account of the misses, a heavy upper tail that the nine Phase~1 cells share, is NOT SUPPORTED, because it needs all four of its parts. 
Three parts hold (HELD), among them $\kappa>0$ (Section~\ref{sec:attacks}) in 8 of the 9 Phase~1 cells. 
The fourth fails: across the nine cells the Spearman correlation between $\kappa$ and the mean $\log$ of realised over predicted rate at $c<0$ is $-0.367$ (FAILED), a quantity dominated by the resolution floor, and no A1 verdict changes. 

\textit{Poisoning the realignment.}
We predicted that the high-score sets of Section~\ref{sec:res-repair}, injected into the realignment's reference pool at a share $\rho\le1\%$, would move the realigned false-positive rate by $\times2$ or more in at least 80\% of the cells. 
The prediction FAILED (P-RA.4) in all six arms, the ridge and the orthogonal Procrustes map on Phase~1, Phase~2 and LoRA. 
The realigned rate moved by $\times2$ or more in 5 of 162 cells, all of them in Phase~2 with the ridge map (5 of its 27), and in none of the 27 cells of each other arm. 
Realignment is thus far less exposed to this poisoning than we predicted.

\textit{A3.}
Each of A3's four registered predictions, P-A3.1 to P-A3.4, holds where it is defined. 
In 22 of A3's 36 cells its registered interval $(0.5,2)$ for the estimated-rate check admits no attacker, because the check already fires on the clean pool. 
For P-A3.1, 0.935 of its 432 cells (pair, reference with its score repair, attacker and $\rho$; restandardisation on traffic is scored by P-A3.2 and P-A3.4) are within $\times1.5$ (HELD), but in 288 of them, the threshold re-set cells, the predicted and realised true-positive rates match by construction, because the prediction is the re-set's own rule applied to the same pool; of the other 144, the quantile-transfer cells, 0.806 are within $\times1.5$ (descriptive). 
Two sensitivity analyses sit beside A3 and never replace it: poisoning with held-out benign candidates HELD (SB), and the clean-relative band FAILED in one pair, Mistral (SA). 

\textit{A2.}
A2, the topic-targeted update attack, kept overall AUROC, with no cell falling by 0.05 or more (P-A2.1, HELD), and kept the overall false-positive rate within $\times1.5$ of the clean update's in 30 of 30 cells (P-A2.4, HELD). 
Every A2 attacked model stayed inside the KL bound. 
Under the registered rule it succeeded on one category: cat-3 gave a stealthy cell (Section~\ref{sec:metrics}) in 3 of its 5 seeds at $c=-2$. 
That refutes our registered hypothesis that no single attacker evades both the AUC rule and the estimated-rate check (Hypothesis A2, REFUTED). 
Our prediction that every target category would lose 10 points at $c=-2$ FAILED (P-A2.3). 
Seven of 30 cells failed stealth only on the ARC-Easy condition, each by 1 to 4 of 2376 questions. 
One of those seven decides a second category's outcome, which is 2 of 5 stealthy seeds as scored and would be 3 of 5 with that cell counted. 
Cat-3's three stealthy cells sit 4, 2 and 1 questions above the ARC-Easy clean range's minimum of 1626 of 2376, and A2 met its registered rule at the minimum: one category, 3 of 5 seeds. 
The estimated-rate check flagged 7 of 30 cells, more than we predicted (P-A2.2, FAILED); by definition, cat-3's three stealthy cells are not among them. 

\textit{The canary.}
A canary check flagged every blinded cell at $c=-2$ (Section~\ref{sec:metrics}), 8 of 8, or 9 of 9 on the frozen-threshold reading, with no false alarm in 603 clean draws, which its threshold rule guarantees by construction (Section~\ref{sec:attacks}). 
Cat-3's three stealthy cells are among them, and the canary flagged 12 of 30 cells in all. 
The 9 add the one cell at $c=-2$ that lost 10 or more points at the frozen threshold but not at $R_{\text{same}}$. 
Canaries measure the loss only loosely, within $\times1.5$ of the evaluation loss in 5 of 9 cells against the pre-update model a deployer has. 
They protect only the categories they cover, and only while they stay secret from the attacker. 
\space%
\section{Discussion and Limitations}\label{sec:discussion}

\subsection{Implications for deployment}\label{sec:disc-impl}

\textit{Re-measure after an update.}
After an update, a deployer should re-measure the false-positive rate at $T$ by scoring stored benign prompts through the updated model, which needs no new labels (Section~\ref{sec:setting}). 
Concurrent work makes the same recommendation~\cite{anon2026basecase}.
Estimating the rate finds far more of the damage than the AUC rule does, and re-measuring on the stored prompts finds about as much, but neither is exact: each misclassifies some of the 156 Phase~1 layer cells (Section~\ref{sec:res-mech}). 

\textit{Fragility depends on the target rate.}
Of the 11 headline cells classed fail closed or fail open at 1\%, five are within at 5\% and six keep their class (descriptive; Section~\ref{sec:res-five}). 
A finding at one target rate therefore need not hold at another, so a deployer should re-measure at the rate in use, and a looser target does not remove the need to re-measure.
The pre-update dispersion test also detects damage in more Aegis pairs at 5\% than at 1\%, so the classes do not show that a looser target is safer in general. 

\textit{Re-setting the threshold.}
The re-set's cost is one pass of the 1000 stored calibration negatives through the updated model and a new $T$ computed from their scores; the probe, its standardisation and $L^*$ stay as deployed. 
It needs no new labels, but it needs the stored prompts themselves, not only their pre-update scores, and the updated model's activations at $L^*$.
It needs benign prompts from the population on which $T$ was set: our stored and fresh references are drawn from that population (Section~\ref{sec:repairs}), and concurrent work finds that matching on traffic from another population leaves the rate off target~\cite{anon2026basecase}.
It also needs a reference that no attacker can write to (Section~\ref{sec:res-repair}).
It restores the false-positive rate, not the pre-update true-positive rate (Section~\ref{sec:res-repair}).

\textit{What realignment buys.}
Realignment keeps AUROC by a criterion we extended from Duan's, with either map form, but in our cells the frozen probe already met that criterion, so by it there was nothing to repair (Section~\ref{sec:res-repair}). 
Neither map form reliably returns the false-positive rate at $T$ to its target, so a deployer who realigns must still measure that rate. 
Realignment was also far less exposed to poisoning of its reference than we predicted (FAILED): the only cells whose realigned rate moved by $\times2$ or more were ridge cells in Phase~2 (Section~\ref{sec:res-attack}). 

\subsection{Limitations}\label{sec:disc-lim}

Our design leaves four alternative explanations open.
Phase~1's pairs were chosen by no registered rule and 8 of their 9 headline cells fail closed at 1\%; Phase~2's were, and 3 of 9 fail closed or fail open (Table~\ref{tab:headline}). 
Phase~1 also reads each instruct model on the raw prompt, not through its chat template, so the read-out, the kind of update and the phase are aliased (Section~\ref{sec:res-op}). 
Pairs 2b, 2c and 2e change the rendered input with the weights. 
Two of them, 2b and 2e, fail closed at 1\% (Table~\ref{tab:headline}). 
Under the registered sensitivity pooling at the last templated position, those three cells change class: 2b and 2c fail open, 2e is within, and no Phase~2 headline cell fails closed, against two under our pooling. 
That arm also re-selects $L^*$ and $C$, so it does not isolate the pooling from the layer, and 2e's AUROC change there reaches $-0.0478$, against $-0.05$ for a big drop. 
The comparison is descriptive: no registered prediction scores this pooling (Supplement~\ref{supp:beside}). 
The updates 1e, 1g, 2a, 2b and 2d also change the model configuration, as do 1h and 2j outside the headline set (Table~\ref{tab:config}), and we do not separate these changes from the change in weights. 
Of these seven pairs, five are version increments or instruct-side releases and 2j changes the tokeniser, while only one of our six base-to-instruct pairs, 1e, is among them; this is why, for our 18 numbered pairs, we define an update as the released weights together with their released configuration and, in Phase~2, their released format. 
The HH-harmless arm's exposure remains, and no current record bounds paraphrase or near-duplicate contamination of the benign set by the fine-tunes' training data (Section~\ref{sec:res-lora}). 

For four headline cells, 1a and 1c on Aegis~2.0, 1e and 1g, some input caches have no extraction record, so the same GPU type on both sides rests on a note in the record that verified those caches, which we read post hoc (Supplement~\ref{supp:provenance}). 
Three of the four, all but 1e, are among the 11 cells classed fail closed or fail open (Table~\ref{tab:headline}). 
The hardware control changes the GPU and keeps the weights, and it bounds the effect of that change on the paired ratio, but we ran it on Mistral-7B v0.1 alone (Section~\ref{sec:pairs}). 
The phase with most of the damage is also the phase with these gaps: Phase~1 holds 8 of the 10 cells that fail closed at 1\%, its pairs follow no registered rule, and all four cells above are Phase~1 cells, among them 1a's Aegis~2.0 cell, the largest rise. 
Of the five Phase~1 cells whose input caches all have extraction records, 1b, 1d, 1f and the WildGuardMix cells of 1a and 1c, all five fail closed on the mean paired ratio, while on the band three lie wholly above the cut at 2 and two, 1c's and 1d's, contain it. 

Only one headline cell, 2a, fails open, and nothing general follows from one cell: it shows neither that updates tend to fail open nor that they rarely do. 

We studied only open-weight models of up to 9B parameters. 
The controlled LoRA updates use one checkpoint, Mistral-7B v0.1, and the model-side attacks one backbone, Mistral-7B-Instruct-v0.2. 
The probes are trained on two safety datasets, Aegis~2.0 and WildGuardMix, and over-flagging is measured on one set, XSTest. 
Adaptive attackers are out of scope: no attack in this study adapted to the canary, which we fixed after A2 was scored (Section~\ref{sec:res-attack}). 

Our worst frozen-probe change at $L^*$ under our pooling, $-0.0470$ over the realignment study's draws, clears Duan's big-drop line of $-0.05$ by 0.003 of AUROC, so ``no stale cell'' rests on that margin and on the seed mean, since on the fixed split one of that run's five seeds crosses the line (Section~\ref{sec:res-op}); the run is in the HH-harmless arm, whose training data overlap the evaluation items. 
The nine LoRA runs' median change, $-0.0088$ or $-0.0091$ on the two sets of draws (descriptive), is smaller than Duan's LoRA median of $-0.0374$~\cite{duan2026monitors}, pooled over 720 cells, and the layer-averaged mean falls of 5.1 to 9.7 points that Bakker and Louizos~\cite{bakker2026relative} report for standard probes under LoRA. 
Bakker and Louizos use Qwen2.5-3B and Qwen2.5-7B on XSTest with refusal fine-tunes of 256 or 512 examples; we use one Mistral checkpoint on Aegis~2.0. 
No record of ours separates model, data, task or dose, so the difference is unexplained.

Of the 137 entries in our registry, 69 HELD, 24 FAILED, 5 were NOT SUPPORTED, 3 were REFUTED and 6 were withdrawn before running, undefined or inconclusive; the other 30, outcome-only or descriptive, state no prediction (Supplement~\ref{supp:registry}). 
By our post hoc classification, the substantive predictions among them are 52 of the 69 HELD, 22 of the 24 FAILED, all 5 NOT SUPPORTED, all 3 REFUTED and 4 of the other 6; the rest are algebraic identities or anchor and bookkeeping checks. 
By our post hoc reading, at least 16 of the 24 FAILED predictions, among them at least 15 of the 22 substantive ones, failed by predicting more of the effect than the data showed. 
Each prediction that FAILED, was NOT SUPPORTED or was REFUTED is reported in the section where it bears on a claim. 

\subsection{Registration and records}\label{sec:disc-records}
For 83 of the 88 registered predictions that carry a claim the paper states, in the main text or the supplement, the commit precedes the result's recorded run time; for 2 it does not, and we say so where they are used; for the other 3, registered rows~19 to~21, the registry gives the scored values but names no result file, so there is no recorded run time to compare, and their arm's training began about a minute after the registering commit, by the clocks of two different machines; the repository is private, so its server-side push log is available on request rather than to readers. 
Registered entries for three analyses class cells by their mean paired ratios: we wrote two of them after the means they class existed, and the third, for the MLP probe (Section~\ref{sec:res-mlp}), after the linear probe's means but before the MLP probe's own. 
A rule registered earlier on the fixed split would withdraw a pair's damage claim when its frozen arm is at the floor set by its no-update null; it recorded an outcome and no verdict, and we state no outcome because every record it was scored on is superseded under our provenance rule, as the provenance of their code cannot be verified (Supplement~\ref{supp:provenance}). 
\space%
\section{Conclusion}\label{sec:conclusion}
The change that a backbone update makes to a frozen probe's false-positive rate follows the benign scores' first two moments, within a factor of 1.5 in most of the Phase~2 cells whose predictions we registered before any Phase~2 evaluation score existed (HELD), and less closely in the cells whose rate moved. 
At the 1\% target the frozen threshold fails closed in 8 of 9 Phase~1 headline cells and fails closed or fails open in 3 of 9 Phase~2 cells (descriptive), phases that also differ in the kind of update and in how the model reads the prompt (Section~\ref{sec:res-op}). 
AUROC-based staleness checks miss this: one such check, built from a recent benchmark's criteria, calls none of the 18 stale. 
After every update, a deployer should re-measure the false-positive rate at $T$, at the target rate in use, by scoring stored benign prompts through the updated model, which needs no new labels.
If the rate has moved, a deployer can re-set the threshold on the same prompts.
Concurrent work reports such failures and makes both recommendations~\cite{anon2026basecase}.
The stored prompts must be out of an attacker's reach: in the nine Phase~1 cells, injecting the high-score sets as 0.5\% or 1\% of a 1000-item reference costs the re-set 10 points of true-positive rate, and at 1\% the re-set flags 0.006 to 0.017 of harmful prompts (descriptive). 
Nor does the operating point move only by accident: an attacker who controls the update moved it on one of three targeted harm categories, in three of five seeds, while both the AUC rule and the estimated-rate check stayed silent, refuting our registered hypothesis that no single attacker evades both (Hypothesis~A2, REFUTED). 
A post hoc canary on held-out labelled harmful prompts of all three categories flagged every such cell, but we fixed the canary after the attack was scored and ran no adaptive attacker. 
\space%

\ifanonymous\relax\else

  \textit{Data and code availability.}
  The code, the registry of our predictions and the result record behind each reported result are available at \url{https://github.com/ShirePyDev/frozen-probe-operating-point}, except for the canary preview of Supplement~\mbox{\ref{supp:disclosures}}, whose printed counts are in our decision log; the released record of the registered run recomputes them.
  The code is released under the MIT License.
  The frozen dataset splits are released as item identifiers and SHA-256 hashes only; no dataset prompt text is redistributed.
  The cached hidden-state activations and probe scores are not released, because of their size (209~GiB); they can be regenerated from the included code and the checkpoint revision hashes pinned in Supplement~\mbox{\ref{supp:revisions}}. 
  The records name the platform and Python version they ran under (Linux GPU hosts with Python~3.10.12; for some analyses, a Windows workstation with Python~3.12.10), apart from 14 that name none; only those written from 25~September~2026 on also name their package versions, so the records of Phase~1's extraction do not. 
  The LoRA adapters, the attacked models of Section~\ref{sec:res-attack} and the canary prompt sets are not released; the training code is included, so both the adapters and the attacks can be reproduced.
  The datasets are public and are used under their own licences: Aegis~2.0, WildGuardMix (train and test), XSTest (a copy of its v2 release), SST-2, Alpaca (cleaned), HH-RLHF, PKU-SafeRLHF and ARC-Easy. 
\fi

\newcommand{\gccfackheading}{\ifCLASSOPTIONcompsoc\section*{Acknowledgments}\else\section*{Acknowledgment}\fi}
\ifanonymous\else
  \gccfackheading
  This work was supported in part by the National Research Foundation of Korea (NRF) grant funded by the Korea Government [Ministry of Science and ICT (MSIT)] under Grant RS-2026-25522067; and in part by the Institute of Information and Communications Technology Planning and Evaluation (IITP) grants funded by the Korea Government (MSIT) under Grant IITP-2024-RS-2024-00397085 (Leading Generative Artificial Intelligence (AI) Human Resources Development), Grant IITP-2024-RS-2024-00417958 (Global Research Support Program in the Digital Field), and Grant RS-2025-02219317 (AI Star Fellowship, Kookmin University). 

\fi
\ifanonymous
  \gccfackheading
\fi

  The research questions, the claims and the rules under which the work was carried out were the authors', who retained authority over the wording of every claim.
  Claude (Anthropic) was used to generate the code for the experiments and the analysis, to propose designs and analyses and draft the pre-registered predictions, and to draft and revise the text apart from the biographies. 
  The authors reviewed and edited the content as needed and take full responsibility for the content of the publication.

\ifhascite
  \bibliographystyle{IEEEtran}
  \bibliography{IEEEtran_bstctl,references}
\fi

\ifanonymous
\else
  \begin{IEEEbiography}[{\includegraphics[width=1in,height=1.25in,keepaspectratio]{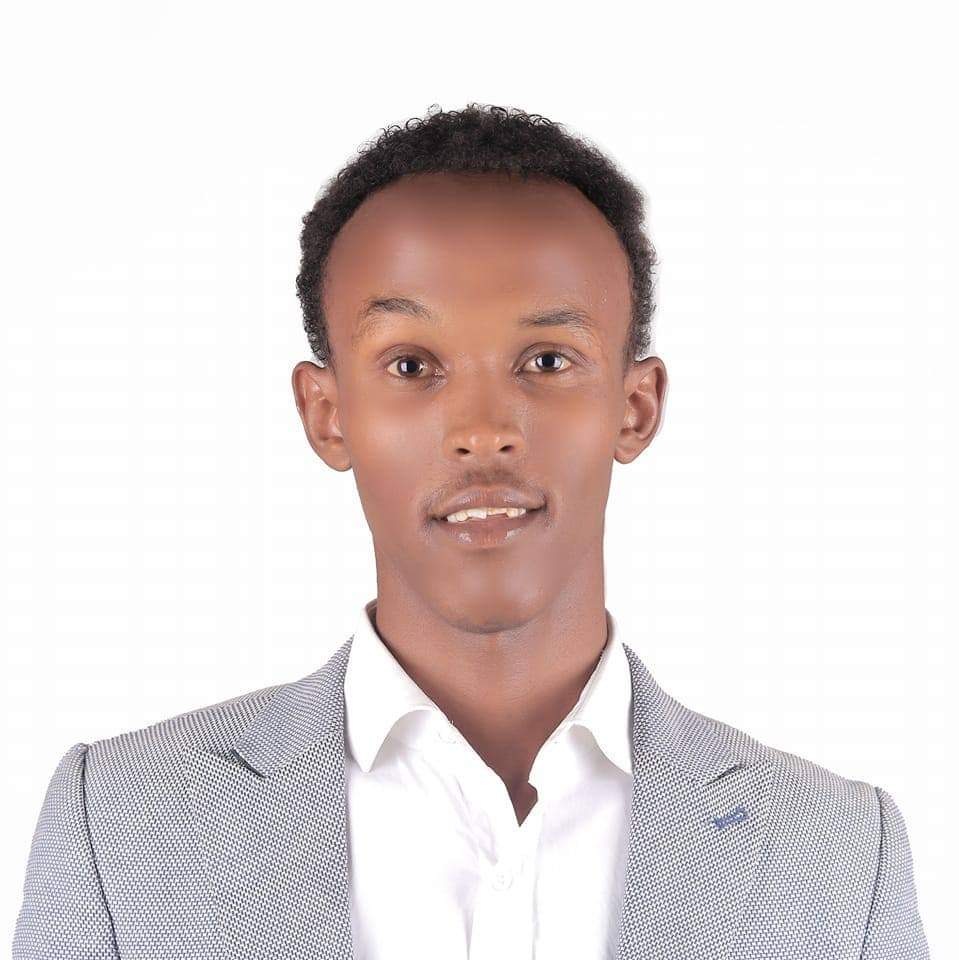}}]{Yusuf Khalid Shire}
    received the B.S. degree in computer science from the University of Bosaso, Bosaso, Somalia, in 2022. He is currently pursuing the M.S. degree in computer engineering, with a major in artificial intelligence convergence, at Kookmin University, Seoul, Republic of Korea, where he is with the Mobile Computing Laboratory. His research interests include the security of large language models, prompt-injection detection, and guardrail design. 
  \end{IEEEbiography}

  \begin{IEEEbiography}[{\includegraphics[width=1in,height=1.25in,keepaspectratio]{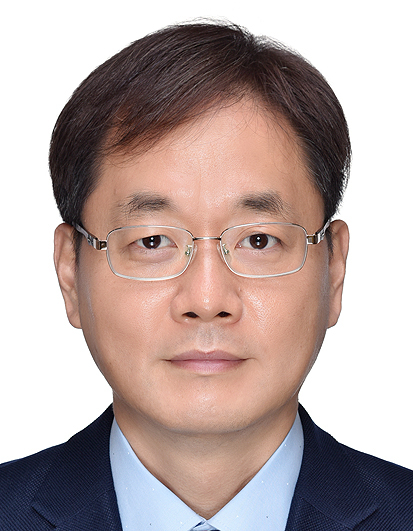}}]{Sang-Chul Kim}
  received the Ph.D. degree in computer engineering from Oklahoma State University, Stillwater, OK, USA, in 2005. 
  He was a Software Engineer with Samsung SDS and Samsung Aerospace from 1994 to 2000. 
  Since 2006, he has been with the School of Computer Science, Kookmin University, Seoul, Republic of Korea, where he is currently a Professor. 
  His research interests include AI security, adversarial robustness of large language and vision-language models, secure edge AI, and software-defined networking.
  He is a Fellow of the Korean Institute of Communications and Information Sciences (KICS) and serves as an Area Editor of KSII Transactions on Internet and Information Systems.
  \end{IEEEbiography}
\fi

\includepdf[pages=-,noautoscale]{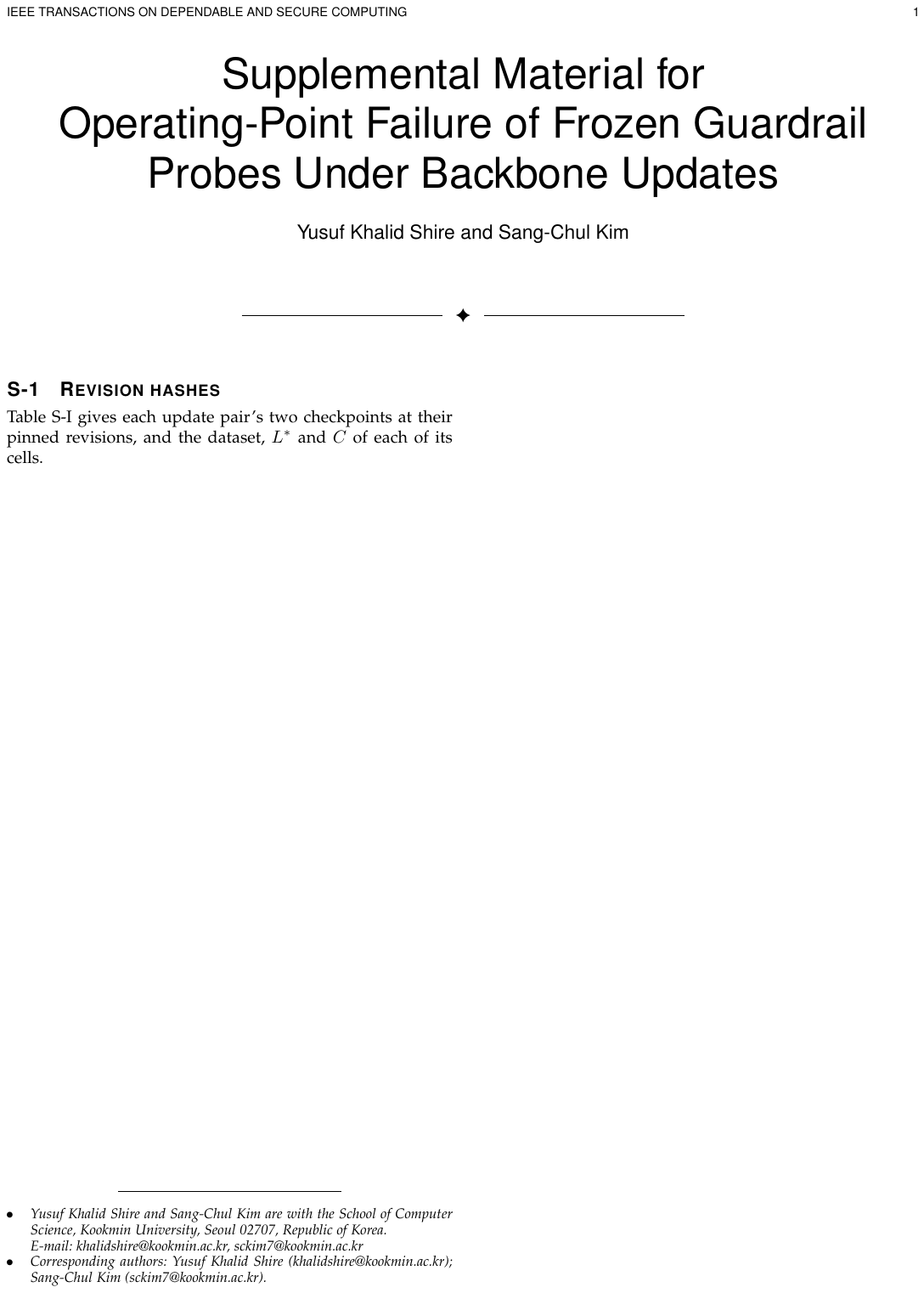}
\end{document}